\documentclass[twocolumn,tighten]{aastex62}

\usepackage{graphicx}
\usepackage{ulem}
\usepackage{amsmath}
\usepackage{wrapfig}

\usepackage{tikz}

\definecolor{tiffany}{RGB}{79, 166, 158}

\newcommand\msun{\, \rm M_\odot}

\newcommand\lsim{\mathrel{\rlap{\lower4pt\hbox{\hskip1pt$\sim$}}
        \raise1pt\hbox{$<$}}}
\newcommand\gsim{\mathrel{\rlap{\lower4pt\hbox{\hskip1pt$\sim$}}
        \raise1pt\hbox{$>$}}}

\begin{document}
\shorttitle{Formation of massive black holes}
\shortauthors{Kremer et al.}

\title{Populating the upper black hole mass gap through stellar collisions in young star clusters}

\correspondingauthor{Kyle Kremer}
\email{kkremer@caltech.edu}

\author[0000-0002-4086-3180]{Kyle Kremer}
\altaffiliation{NSF Astronomy \& Astrophysics Postdoctoral Fellow}
\affil{Center for Interdisciplinary Exploration \& Research in Astrophysics (CIERA) and Department of Physics \& Astronomy, Northwestern University, Evanston, IL 60208, USA}
\affiliation{TAPIR, California Institute of Technology, Pasadena, CA 91125, USA}
\affiliation{The Observatories of the Carnegie Institution for Science, Pasadena, CA 91101, USA}

\author[0000-0003-0930-6930]{Mario Spera}
\affil{Center for Interdisciplinary Exploration \& Research in Astrophysics (CIERA) and Department of Physics \& Astronomy, Northwestern University, Evanston, IL 60208, USA}
\affil{Dipartimento di Fisica e Astronomia `G. Galilei', University of Padova, Vicolo dell'Osservatorio 3, I--35122, Padova, Italy}
\affil{INFN, Sezione di Padova, Via Marzolo 8, I--35131, Padova, Italy}

\author{Devin Becker}
\affil{Department of Physics and Astrophysics, DePaul University, Chicago, IL 60614, USA}
\affil{Center for Interdisciplinary Exploration \& Research in Astrophysics (CIERA) and Department of Physics \& Astronomy, Northwestern University, Evanston, IL 60208, USA}

\author[0000-0002-3680-2684]{Sourav Chatterjee}
\affil{Tata Institute of Fundamental Research, Homi Bhabha Road, Mumbai 400005, India}

\author[0000-0003-2654-5239]{Ugo N. Di Carlo}
\affil{Universit\`a degli Studi dell'Insubria, Dipartimento di Scienza e Alta Tecnologia, Via Valleggio 11, I--22100, Como, Italy}
\affil{INFN, Sezione di Padova, Via Marzolo 8, I--35131, Padova, Italy}
\affil{INAF, Sezione di Padova, Vicolo dell'Osservatorio 5, I--35122, Padova, Italy}

\author[0000-0002-7330-027X]{Giacomo Fragione}
\affil{Center for Interdisciplinary Exploration \& Research in Astrophysics (CIERA) and Department of Physics \& Astronomy, Northwestern University, Evanston, IL 60208, USA}

\author[0000-0003-4175-8881]{Carl L. Rodriguez}
\affil{Harvard Institute for Theory and Computation, 60 Garden St, Cambridge, MA 02138, USA}

\author[0000-0001-9582-881X]{Claire S. Ye}
\affil{Center for Interdisciplinary Exploration \& Research in Astrophysics (CIERA) and Department of Physics \& Astronomy, Northwestern University, Evanston, IL 60208, USA}

\author[0000-0002-7132-418X]{Frederic A. Rasio}
\affil{Center for Interdisciplinary Exploration \& Research in Astrophysics (CIERA) and Department of Physics \& Astronomy, Northwestern University, Evanston, IL 60208, USA}

\begin{abstract}

Theoretical modeling of massive stars predicts a gap in the black hole (BH) mass function above $\sim 40-50\,M_{\odot}$ for BHs formed through single star evolution, arising from (pulsational) pair-instability supernovae. However, in dense star clusters, dynamical channels may exist that allow construction of BHs with masses in excess of those allowed from single star evolution. The detection of BHs in this so-called ``upper-mass gap'' would provide strong evidence for the dynamical processing of BHs prior to their eventual merger. Here, we explore in detail the formation of BHs with masses within or above the pair-instability gap through collisions of young massive stars in dense star clusters. We run a suite of 68 independent cluster simulations, exploring a variety of physical assumptions pertaining to growth through stellar collisions, including primordial cluster mass segregation and the efficiency of envelope stripping during collisions. 
We find that as many as $\sim20\%$ of all BH progenitors undergo one or more collisions prior to stellar collapse and up to $\sim1\%$ of all BHs reside within or above the pair-instability gap through the effects of these collisions. We show that these BHs readily go on to merge with other BHs in the cluster, creating a population of massive BH mergers at a rate that may compete with the ``multiple-generation'' merger channel described in other analyses. This has clear relevance for the formation of very massive BH binaries as recently detected by LIGO/Virgo in GW190521. Finally, we describe how stellar collisions in clusters may provide a unique
pathway to pair-instability supernovae and briefly discuss the expected rate of these events and other electromagnetic transients.

\vspace{1cm}
\end{abstract}

\section{Introduction}
\label{sec:intro}

The mass spectrum of stellar-mass black holes (BHs) is among the most hotly debated topics in modern astrophysics. This is driven in large part by the growing catalog of binary BH mergers detected as gravitational-wave (GW) sources by LIGO/Virgo \citep{LIGO2018b} over the past few years, which have complemented earlier constraints upon BH masses obtained from observations of X-ray binaries \citep[e.g.,][]{Bailyn1998,Ozel2010,Farr2011,Corral-Santana2016}. Over the coming years and decades, current (LIGO, Virgo, KAGRA) and future (e.g., LISA, Einstein Telescope, DECIGO) GW detectors promise to provide unprecedented constraints upon the BH mass distribution. Thus, it is essential to advance our theoretical understanding of the various pathways through which stellar BHs may form.

Stellar-mass BHs are expected to form as the end products of the evolution of massive stars, with the final mass of the BH determined by two primary elements, both ripe with uncertainty: the mass of the progenitor star (and core-to-envelope mass ratio) just before core collapse, and the details of the subsequent supernova (SN) explosion. The pre-explosion progenitor mass depends crucially upon (metallicity-dependent) stellar winds \citep[e.g.,][]{Vink2001}. In regard to the SN explosion, a number of theoretical models have been proposed and implemented in various studies which yield varying effects upon the BH mass spectrum \citep[e.g.,][]{Heger2003,Woosley2007,Mapelli2010,Belczynski2010,O'Conner2011,Fryer2012,Spera2015,Belczynski2016b,Sukhbold2016,Ertl2016,Farmer2019}.

Specifically, stars with helium cores in the mass range $\sim
45-135\,M_{\odot}$ are expected to undergo
so-called pair instability SNe (PISNe): after the onset of carbon burning the production of electron--positron pairs leads to a rapid loss of pressure and core contraction. This contraction triggers explosive burning of heavier elements leading to a runaway thermonuclear explosion \citep[e.g.,][]{Fowler1964,Rakavy1967,Barkat1967,Fraley1968}. Stars with helium cores in the range $\sim 65-135\,M_{\odot}$ are thought to be completely destroyed by the PISN, leaving no remnant \citep[e.g.,][]{Bond1984,Fryer2001b,Chatzopoulos2012}. On the other hand, if a star builds a helium core in the range $\sim 45-65\,M_{\odot}$, the pair instability is expected to trigger a series of strong pulsations which efficiently reduce mass and entropy of helium and heavy elements until the pulsing activity has damped. The latter process is known as pulsational pair-instability supernova \citep[PPSN; e.g.,][]{HegerWoosley2002,Woosley2007,Woosley2017,Woosley2019}.

PISNe and PPSNe have a strong imprint upon the BH mass spectrum, as these processes are expected to yield a prominent ``gap'' in BH mass between roughly $40$ and $120\,M_{\odot}$ for BHs formed through single star evolution \citep[e.g.,][]{Belczynski2016b,Spera2017}. While the upper and lower boundaries of this mass gap are uncertain and depend upon various assumptions concerning the evolution of massive stars \citep[e.g.,][]{Belczynski2016b,Woosley2017,Spera2017,Giacobbo2018,limongi2018,Marchant2019,Mapelli2019,Stevenson2019,Farmer2019,bel2020,renzo2020}, studies of the first few GW events seems to corroborate, in general, the theoretical predictions of this mass gap \citep{FishbachHolz2017,TalbotThrane2018,LIGO2018b}.o 

A number of recent studies have shown that formation channels outside the standard single star evolution pathway may in fact provide pathways for populating this mass gap. For example, primordial BHs formed through collapse of gravitational instabilities in the early universe may not be subject to the same constraints as BHs formed through stellar evolution and may therefore occupy the mass gap \citep[e.g.,][]{Carr2016}. Additionally, mass-gap BHs may be born through the merger of two smaller BHs \citep[e.g.,][]{MillerHamilton2002,McKernan2012,Rodriguez2018b,Rodriguez2018a,Antonini2019,GerosaBerti2019,McKernan2019}.

Alternatively, heavy BHs may also be formed from the collapse of anomalously massive progenitor stars that form via stellar collisions or mergers of massive binaries. These collisions may occur particularly often in dense stellar environments such as young star clusters \citep[e.g.,][]{PortegiesZwart2004,Gurkan2006,Giersz2015,Mapelli2016}. Such ``collisional runaway'' episodes have traditionally been touted as a formation channel for the elusive class of intermediate-mass BHs (IMBHs) with masses in the range $\sim10^2-10^4\,M_{\odot}$
\citep[see, e.g.,][for a review]{Greene2019}. More recently, \citet{Spera2019,DiCarlo2019,DiCarlo2020,Banerjee2020} revisited this topic in the specific context of the pair-instability mass gap and showed that stellar collisions/mergers in young clusters may also provide a viable pathway for creating BHs with masses forbidden by single star evolution.

The potential role of star-cluster dynamics in creating BHs with masses within and/or above the pair-instability gap is particularly noteworthy. Over the past decade, a growing number of stellar-mass BH candidates have been identified in the Milky Way globular clusters (GCs) through both X-ray/radio \citep{Strader2012,Chomiuk2013,Miller-Jones2015,Shishkovsky2018} and dynamical measurements \citep{Giesers2018,Giesers2019}, suggesting that at least some GCs retain populations of BHs at present \citep{Kremer2019a,Weatherford2018,Weatherford2019}. This observational evidence has been complemented by recent computational simulations of GCs which have demonstrated that realistic clusters can naturally retain hundreds to thousands of BHs throughout their complete lifetimes \citep[e.g.,][]{Morscher2015,Kremer2020}. Additionally, it is now clear that BH populations play a significant role in shaping the long-term dynamical evolution and present-day structure of GCs \citep{Merritt2004,Mackey2007,Mackey2008,BreenHeggie2013,Peuten2016,Wang2016,Chatterjee2017a,ArcaSedda2018,Kremer2018b,Kremer2019a,Zocchi2019,Antonini2019,Kremer2020}.

Furthermore, the dynamical processes relevant to BHs in stellar clusters have emerged as a viable formation channel for binary BH mergers similar to those detected to date by LIGO/Virgo \citep[e.g.,][]{PortegiesZwart2000,O'Leary2009,Banerjee2010,Rodriguez2015a, Rodriguez2016a,AntoniniRasio2016,Askar2017, Chatterjee2017b,Chatterjee2017a, Hoang2018,Samsing2018c,Fragione2018b, Zevin2018}.\footnote{In addition to dynamical formation in dense star clusters, a number of other binary BH formation channels have been proposed, including isolated massive binary evolution \citep[e.g.,][]{Dominik2012,Dominik2013,Belczynski2016a,Belczynski2016b}, GW capture of primordial BHs \citep[e.g.,][]{Bird2016,Sasaki2016}, secular interactions in hierarchical triple systems \citep[e.g.,][]{AntoniniRasio2016,Antonini2017,Silsbee2017,Liu2017,Hoang2018,Leigh2018,FragioneGrishin2019,Fragkoc2019,Fragleipern2019,RodriguezAntonini2018}, and active galactic nuclei disks \citep[e.g.,][]{McKernan2012,Secunda2019,Yang2019,McKernan2019}.} 
Mergers involving BHs in the mass gap would have properties difficult (or impossible) to produce through isolated binary evolution. Thus, the detection of binary BH mergers with component masses within the pair-instability mass gap \citep[for example the recent event GW190521;][]{LIGO2020a,LIGO2020b} may be strong evidence for the dynamical processing of BHs prior to their eventual merger.

In this analysis, we investigate in detail the formation of massive BHs through stellar collisions in dense star clusters. \citet{DiCarlo2019,DiCarlo2020} explored this topic in the context of lower mass clusters ($\approx10^3-10^4\,M_{\odot}$) and found that, depending on the assumed metallicity, as many as $\sim5\%$ of all BHs in young star clusters can have masses in the pair-instability gap as a result of collisions of young massive stars. They also showed that mass-gap BHs efficiently acquire BH binary companions and merge through subsequent dynamical encounters, yielding a subpopulation of BBH mergers detectable by LIGO/Virgo with at least one component in the mass gap.

Here, we examine the role of stellar collisions on BH formation in the previously unexplored massive cluster regime, exploring specifically clusters with masses comparable to the GCs observed in the Milky Way \citep[$\approx 10^5-10^6\,M_{\odot}$;][]{Harris1996}. This is a critical addition to previous literature on the topic, given that GCs and their progenitors may account for a large fraction of the overall BBH merger rate in the local universe \citep[e.g.,][]{RodriguezLoeb2018,Kremer2020}.

We describe our computational methods for modeling dense star clusters in Section \ref{sec:method}. In Section \ref{sec:results}, we discuss the results of our simulations, describing specifically the various evolutionary outcomes for massive stars undergoing collisions in star clusters. We also discuss the long-term fate of the massive BHs that form through collisions, in particular investigating the possibility of BBH mergers. In Section \ref{sec:GW}, we discuss implications for GW astronomy, in particular comparing to the recently detected upper-mass gap event GW190521.
In Section \ref{sec:PISN}, we discuss briefly the cosmological rates of PISNe and other electromagnetic transients identified in our cluster simulations. We discuss our results and conclude in Section \ref{sec:conclusion}.

\section{Modeling Cluster Evolution}
\label{sec:method}

We use \texttt{CMC} (for \texttt{Cluster Monte Carlo}) to model the evolution of stellar clusters. \texttt{CMC} is a distributed-memory parallelized H\'{e}non-type Monte Carlo code that includes prescriptions for various physical processes relevant to the problem at hand including two-body relaxation, up-to-date stellar/binary evolution from the population synthesis code \texttt{COSMIC} \citep{Breivik2020}, direct integration of small-$N$ resonant encounters \citep[using the \texttt{fewbody} package;][]{Fregeau2004}, tidal mass loss \citep{Chatterjee2010}, and stellar collisions \citep{Fregeau2007}. For a recent review of the computational method of \texttt{CMC}, see \citet{Kremer2020} and references therein. Here, we make several changes to \texttt{CMC} to explore the particular subject of massive star collisions and implications for the formation of massive BHs. We summarize these changes below:

\textit{Primordial mass segregation:} Observations of many young massive clusters (YMCs) show an increased concentration of massive stars near the cluster centers \citep[e.g.,][]{Hillenbrand1997,Hillenbrand1998,Fischer1998,Gouliermis2004,Stolte2006}. The origin of this mass segregation is uncertain. Mass segregation is known to be a natural feature of self-gravitating systems driven by two-body relaxation \citep[e.g.,][]{Spitzer1987,HeggieHut2003}. However, mass segregation is observed in many YMCs with ages much less than their relaxation times, suggesting that it may in fact be a primordial feature of at least some clusters. Primordial mass segregation has been proposed to result, for example, from the preferential formation of massive stars in the densest regions of the molecular cloud \citep[e.g.,][]{MurrayLin1996} or by gas accretion during the initial phases of star formation \citep{BonnellBate2006}.

\begin{figure}
\begin{center}
\includegraphics[width=0.95\linewidth]{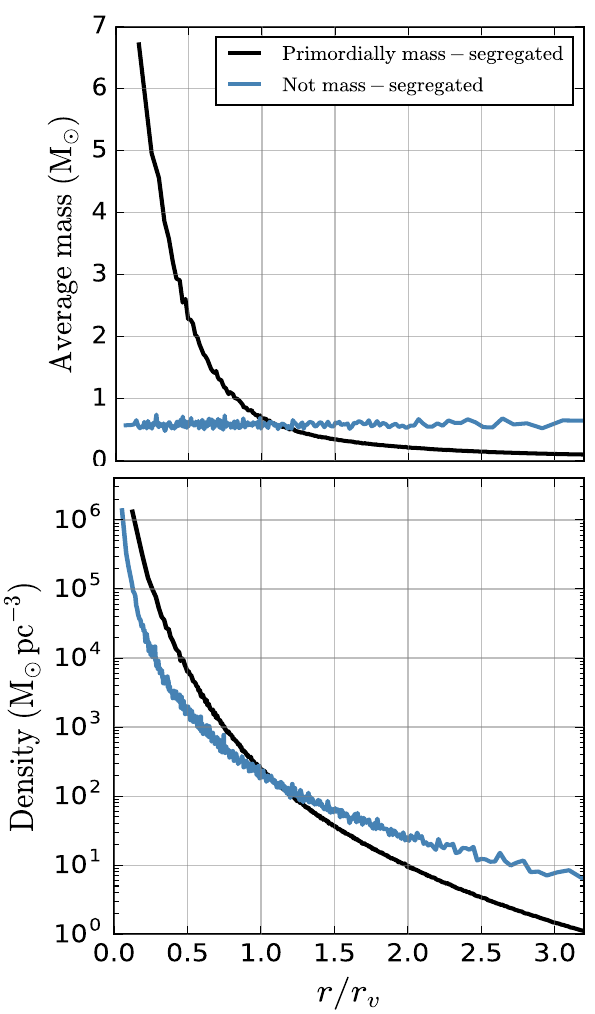}
\caption{\footnotesize \label{fig:avg_mass} \textit{Top panel:} Average stellar mass profile versus radius (in units of initial virial radius) for a non-mass-segregated cluster (blue) and a primordially mass-segregated cluster (black) initialized using the recipe of \citet{Baumgardt2008}. \textit{Bottom panel:} Density profile for the same two models.
}
\end{center}
\end{figure}

A number of recent studies have explored the effect of primordial mass segregation on the formation of massive stars through collisional runaway \citep[e.g.,][]{Gurkan2004,Ardi2008,Goswami2012}. Such runaways have been proposed as a potential formation channel for IMBHs \citep[e.g.,][]{PortegiesZwart2002,Freitag2006}. To investigate these potential effects, we run simulations both with and without primordial mass segregation. We adopt the recipe of \citet{Baumgardt2008} to create primordially mass-segregated clusters in virial equilibrium. In this recipe, stars are sorted such that (for a fixed number density profile) the most massive stars have, on average, the lowest specific energy. For illustration, in Figure \ref{fig:avg_mass}, we show the average stellar mass profile (top panel) and mass density profile (bottom panel) for two clusters with identical particle numbers and initial mass functions with and without primordial mass segregation. Here we show radial position in units of the cluster virial radius, $r_v=GM_c^2/(2U)$, where $M_c$ is the cluster mass and $U$ is the total cluster potential energy. See also \citet{Goswami2012} for a recent detailed examination of the effects of primordial mass segregation using \texttt{CMC}.

\vspace{0.25cm}

\textit{Initial stellar binaries and three-body binary formation:} In order to isolate
the effect of dynamical collisions on stellar growth, we assume zero stellar binaries in all simulations in this study.
As a result, all stellar collisions considered in this study occur through single--single encounters in which the pericenter distance of the pair of stars at closest approach is less than the sum of radii of the pair ($r_p \leq R_1+R_2$). See \citet{Fregeau2007} for a detailed explanation of the treatment of single--single collisions in \texttt{CMC}.
Recent analyses have shown that binary evolution processes (e.g., mass transfer, common envelope, tides) can play in important role in the formation of massive BHs both in the field and in star clusters \citep[e.g.,][]{Spera2019,DiCarlo2019,DiCarlo2020}. Thus, by not including stellar binaries, the results of this study may be viewed as a lower limit on the formation rate of massive BHs in dense star clusters.

Although we do not include stellar binaries in our models, once BHs form, we do allow BH binaries to form through three-body binary formation \citep[e.g.,][]{HeggieHut2003}. We follow the formalism outlined in \citet{Morscher2013}, allowing BH binaries to form with $\eta \geq 2 = \eta_{\rm{min}}$, where $\eta$ is the binary hardness ratio (binary binding energy to background star kinetic energy). The formation of BH binaries through this mechanism is essential for the ejection of BHs throughout the cluster evolution and also the formation of GW sources \citep[see][for a review]{Kremer2020}.

\vspace{0.25cm}

\textit{Treatment of stellar collision products:} The ultimate fate of a stellar collision product depends upon the details of the collision, which in turn depend on the internal structures of the two colliding stars.
For collisions of two main sequence stars, we adopt the so-called ``sticky sphere'' approximation where we assume no mass is lost during the collision itself such that $M_3$, the mass of the collision product, is simply equal to the sum of the masses of the two colliding stars, $M_1+M_2$. The stellar age of the new main sequence star is given by 

\begin{equation}
    t_3 = f_{\rm{rejuv}} \frac{t_{\rm{MS3}}}{M_3} \Bigg( \frac{M_1 t_1}{t_{\rm{MS1}}} + \frac{M_2 t_2}{t_{\rm{MS2}}} \Bigg)
\end{equation}
where $t_{\rm{MS1}}$, $t_{\rm{MS2}}$, and $t_{\rm{MS3}}$ are the MS lifetimes of the two collision components and the collision product, respectively, and $t_1$ and $t_2$ are the stellar ages of the two collision components at the time of collision. $f_{\rm{rejuv}}$ is a factor which determines the amount of rejuvenation the collision product experiences through mixing. In reality, this factor depends upon the internal structure of the two stars as well as the nature of the collision (i.e., the impact parameter and velocity at infinity). In original \texttt{BSE} \citep{Hurley2002}, a fixed value of 0.1 is assumed for $f_{\rm{rejuv}}$. However, in many instances this likely leads to over-rejuvenation of the collision product. Therefore we adopt $f_{\rm{rejuv}}=1$ as our default value \citep[see also][for discussion]{Breivik2020}.

If on the other hand, at least one of the collision components is a giant, the complete mixing scenario relevant for MS--MS collisions is no longer applicable and instead the collision is qualitatively more similar to a common envelope episode where the cores of the two stars inspiral within an envelope of more loosely bound material. In this case, it is less clear whether the sticky sphere assumption is appropriate. Therefore, we adopt two different prescriptions for collisions involving giant stars that bracket the range of expected outcomes. As an upper-limit case, we assume sticky sphere collisions where the collision product's total mass is again computed as $M_3=M_1+M_2$ and the core mass is computed as $M_{c,3}=M_{c,1}+M_{c,2}$ (note that we assume $M_c=0$ for main sequence stars). As a lower-limit case, we assume that the envelopes of giant stars are completely unbound through the collision process. In this limit, in the case of a giant--main sequence collision, $M_3=M_{c,1}+M_2$ and $M_{c,3}=M_{c,1}$, such that the collision product is a giant. In the case of a giant--giant collision, $M_3=M_{c,3}=M_{c,1}+M_{c,2}$ such that the collision product is a naked helium core.\footnote{A naked helium star (stellar type $k=7$) is defined in standard \texttt{BSE} to have $M_c=0$ \citep{Hurley2002}. If (in the lower-limit case where we assume giant envelopes are ejected) a helium star (with mass $M_1$) undergoes a subsequent collision with a giant of mass $M_2$ and core mass $M_{c,2}$, we assume a new naked helium star is formed with total mass $M_3=M_{1}+M_{c,2}$ and core mass $M_{c,3}=0$.  We acknowledge that this simple treatment may miss subtleties associated with such collisions. However, because the radius of a naked helium star is small and the lifetime short, such collisions are rare, thus a more detailed treatment will not have a significant effect upon our results.} 

\vspace{0.25cm}

\textit{Collisional runaways:} Clusters with sufficiently high initial densities may lead to high stellar collision rates and potentially to the formation of a very massive star ($M \gtrsim 1000\,M_{\odot}$) within the first few Myr before the stars undergo core-collapse supernovae. A number of analyses have shown that these very massive stars may have important implications for the formation of IMBHs \citep[e.g.,][]{Ebisuzaki2001,PortegiesZwartMcMillan2002,Gurkan2004,Freitag2006,PortegiesZwart2010,Goswami2012}. For clusters that are primordially mass-segregated, the onset of these collisional runaways may be even more likely \citep[e.g.,][]{Goswami2012}.

Treatment of the various physical processes relevant in the presence of a very massive star (and/or an IMBH) is, at present, beyond the computational scope of \texttt{CMC}. Therefore, if a star of mass $\geq1000\,M_{\odot}$ forms through stellar collisions, we simply stop the simulation and record the outcome as a collisional runaway. Note that this assumed $1000\,M_{\odot}$ threshold is chosen to be roughly consistent with \citet{PortegiesZwart2002}, which showed that the products of collisional runaways can reach masses of up to roughly $0.1\%$ of the total cluster mass.

\vspace{0.25cm}

\textit{Compact object formation:} We adopt the (metallicity-dependent) stellar wind prescriptions of \citet{Vink2001} to determine the final stellar mass at the moment of core collapse. We use the ``delayed'' SNe explosion models \citep{Fryer2012} to compute neutron star and BH masses modified to include prescriptions for PPSNe and PISNe. Following \citet{Belczynski2016b}, we assume that any star with a pre-explosion helium core mass in the range $45-65\,M_{\odot}$ will undergo pulsations that eject large amounts of the hydrogen envelope such that the final stellar mass at the time of core collapse is $45\,M_{\odot}$.
In this case, stars that undergo PPSNe are assumed to yield BHs of mass $40.5\,M_{\odot}$ (we assume that $10\%$ of the final core mass is lost through the conversion of baryonic matter to gravitational matter at the moment of collapse, such that the final remnant mass is $90\%$ of the pre-explosion core mass). We assume stars with pre-explosion core masses in the range $65-135\,M_{\odot}$ undergo PISNe and no compact remnant is formed. Stars with core masses in excess of $135\,M_{\odot}$ are assumed to undergo direct collapse to a BH, such that the BH mass is equal to $90\%$ of the pre-explosion \textit{total} stellar mass, again accounting for $10\%$ mass loss in conversion from baryonic to gravitational matter.\footnote{We note that the stellar evolution of very massive stars may be quite different from that of lower mass stars \citep[e.g.,][]{Chen2015}. Thus, a more detailed study may implement alternative prescriptions for massive star evolution.}

BH and NS natal kicks are computed as in \citet{Kremer2020}. We assume all BHs are born with zero spin \citep[dimensionless spin parameter $a=0$;][]{FullerMa2019} and also assume that BHs can be spun up only through mergers with other BHs (although see Section \ref{sec:conclusion} for discussion of alternative possibilities). In the event of binary BH merger, we compute the spin (as well as mass and GW recoil kick) of the new BH using the method described in \citet{Rodriguez2018}, which in turn implements phenomenological fits to numerical and analytic relativity calculations \citep{Barausse2009,Campanelli2007,Gonzalez2007,Lousto2008,Lousto2012,Lousto2013,GerosaKesden2016}.

\vspace{0.25cm}

In all models we assume $N=8\times10^5$ stars at birth with masses drawn from an initial mass function ranging from $0.08-150\,M_{\odot}$ with slopes following \citet{Kroupa2001}. We assume a metallicity of $Z=0.002$ ($0.1Z_{\odot}$) and adopt a fixed galactocentric distance of $20\,$kpc in a Milky Way-like galactic potential. In order to explore the effect of initial cluster density upon the stellar collision process, we vary the initial cluster virial radius: $r_v=0.8,0.9,1,1.1,1.2\,$pc. We turn on and off primordial mass segregation and also explore the upper and lower limit cases for giant collisions as described above.
This yields a grid of $5\times2\times2=20$ simulations. To increase the statistical robustness of our results, we perform 3--5 independent simulations of each set of initial parameters, giving us 68 simulations in total. As we are interested primarily in exploring the imprint of stellar collisions on the BH mass spectrum, we run each simulation for only $30\,$Myr or until a collisional runaway occurs
(the most massive star in the simulation grows to $\geq1000\,M_{\odot}$; see Section \ref{sec:method}). Table \ref{table:models} includes a complete list of all simulations in this study.

\startlongtable
\begin{deluxetable*}{c|c|c|c|ccccccc}
\tabletypesize{\scriptsize}
\tablewidth{0pt}
\tablecaption{List of cluster simulations \label{table:models}}
\tablehead{
	\colhead{Model} &
	\colhead{$r_v\,(\rm{pc})$} &
	\colhead{Prim. MS} &
	\colhead{Giant coll.} &
    \colhead{$N_{\rm{BH}}$} &
    \colhead{$N_{\rm{BH,coll}}$} &
	\colhead{$N_{\rm{PPSN}}$} &
	\colhead{$N_{\rm{PISN}}$} &
    \colhead{ $40.5 < M_{\rm{BH}} < 120 M_{\odot}$} &
    \colhead{$M_{\rm{BH}} > 120 M_{\odot}$} &
    \colhead{Max BH mass}
}
\startdata
\texttt{1a} & 0.8 & y & SS & \multicolumn{7}{c}{Runaway at $t=3.62$ Myr} \\
\texttt{1b} & 0.8 & y & SS &  \multicolumn{7}{c}{Runaway at $t=3.57$ Myr} \\
\texttt{1c} & 0.8 & y & SS & \multicolumn{7}{c}{Runaway at $t=3.71$ Myr} \\
\hline
\texttt{2a} & 0.9 & y & SS & 2217 & 109 & 72 & 4 & 7 & 0 & 94.2 \\
\texttt{2b} & 0.9 & y & SS & 2205 & 102 & 68 & 8 & 3 & 1 & 168.0 \\
\texttt{2c} & 0.9 & y & SS & 2221 & 119 & 66 & 5 & 3 & 1 & 328.1 \\
\texttt{2d} & 0.9 & y & SS & 2232 & 114 & 72 & 3 & 4 & 0 & 72.5 \\
\hline
\texttt{3a} & 1 & y & SS & 2236 & 72 & 78 & 2 & 3 & 0 & 66.1 \\
\texttt{3b} & 1 & y & SS & 2243 & 75 & 75 & 5 & 0 & 0 & 40.5 \\
\texttt{3c} & 1 & y & SS & 2237 & 76 & 70 & 2 & 1 & 2 & 202.4 \\
\texttt{3d} & 1 & y & SS & 2240 & 87 & 78 & 0 & 2 & 0 & 70.1 \\
\texttt{3e} & 1 & y & SS & 2240 & 103 & 79 & 2 & 2 & 0 & 69.6 \\
\hline
\texttt{4a} & 1.1 & y & SS & 2247 & 59 & 74 & 1 & 1 & 0 & 40.7 \\
\texttt{4b} & 1.1 & y & SS & 2255 & 68 & 79 & 0 & 1 & 0 & 41.4 \\
\texttt{4c} & 1.1 & y & SS & 2253 & 59 & 76 & 0 & 1 & 0 & 66.7 \\
\hline
\texttt{5a} & 1.2 & y & SS & 2256 & 42 & 73 & 1 & 1 & 0 & 45.4 \\
\texttt{5b} & 1.2 & y & SS & 2253 & 48 & 75 & 0 & 1 & 0 & 56.0 \\
\texttt{5c} & 1.2 & y & SS & 2243 & 51 & 73 & 4 & 2 & 0 & 85.0 \\
\texttt{5d} & 1.2 & y & SS & 2252 & 49 & 74 & 1 & 0 & 0 & 40.5 \\
\hline
\texttt{6a} & 0.8 & n & SS & 2246 & 382 & 62 & 0 & 3 & 0 & 64.8 \\
\texttt{6b} & 0.8 & n & SS & 2239 & 360 & 65 & 0 & 4 & 1 & 207.7 \\
\texttt{6c} & 0.8 & n & SS & 2222 & 349 & 69 & 3 & 2 & 1 & 230.5 \\
\texttt{6d} & 0.8 & n & SS & 2227 & 368 & 74 & 2 & 1 & 1 & 623.7 \\
\hline
\texttt{7a} & 0.9 & n & SS & 2247 & 286 & 71 & 3 & 2 & 0 & 55.8 \\
\texttt{7b} & 0.9 & n & SS & 2248 & 307 & 73 & 1 & 2 & 0 & 50.4 \\
\texttt{7c} & 0.9 & n & SS & 2256 & 288 & 68 & 2 & 0 & 0 & 40.5 \\
\hline
\texttt{8a} & 1 & n & SS & 2258 & 240 & 72 & 0 & 0 & 0 & 40.5 \\
\texttt{8b} & 1 & n & SS & 2256 & 235 & 68 & 0 & 0 & 0 & 40.5 \\
\texttt{8c} & 1 & n & SS & 2253 & 232 & 74 & 0 & 1 & 0 & 75.7 \\
\texttt{8d} & 1 & n & SS & 2258 & 240 & 74 & 0 & 0 & 0 & 40.5 \\
\hline
\texttt{9a} & 1.1 & n & SS & 2257 & 190 & 75 & 0 & 0 & 0 & 40.5 \\
\texttt{9b} & 1.1 & n & SS & 2257 & 188 & 76 & 0 & 1 & 0 & 53.5 \\
\texttt{9c} & 1.1 & n & SS & 2257 & 175 & 77 & 0 & 2 & 0 & 53.0 \\
\texttt{9d} & 1.1 & n & SS & 2256 & 193 & 75 & 0 & 0 & 0 & 40.5 \\
\hline
\texttt{10a} & 1.2 & n & SS & 2261 & 156 & 76 & 0 & 0 & 0 & 40.5 \\
\texttt{10b} & 1.2 & n & SS & 2256 & 161 & 74 & 1 & 0 & 0 & 40.5 \\
\texttt{10c} & 1.2 & n & SS & 2258 & 141 & 76 & 0 & 0 & 0 & 40.5 \\
\texttt{10d} & 1.2 & n & SS & 2257 & 162 & 76 & 0 & 0 & 0 & 40.5 \\
\hline
\texttt{11a} & 0.8 & y & EE & 2186 & 140 & 38 & 0 & 3 & 0 & 70.0 \\
\texttt{11b} & 0.8 & y & EE & 2202 & 138 & 43 & 0 & 4 & 0 & 67.3 \\
\texttt{11c} & 0.8 & y & EE & 2203 & 147 & 35 & 0 & 2 & 0 & 79.8 \\
\hline
\texttt{12a} & 0.9 & y & EE & 2225 & 115 & 51 & 0 & 2 & 0 & 70.6 \\
\texttt{12b} & 0.9 & y & EE & 2231 & 102 & 59 & 0 & 1 & 0 & 61.0 \\
\texttt{12c} & 0.9 & y & EE & 2213 & 102 & 53 & 0 & 0 & 0 & 40.5 \\
\hline
\texttt{13a} & 1 & y & EE & 2235 & 85 & 60 & 0 & 2 & 0 & 59.8 \\
\texttt{13b} & 1 & y & EE & 2239 & 78 & 60 & 0 & 0 & 0 & 40.5 \\
\texttt{13c} & 1 & y & EE & 2235 & 81 & 55 & 0 & 0 & 0 & 40.5 \\
\hline
\texttt{14a} & 1.1 & y & EE & 2243 & 63 & 63 & 0 & 2 & 0 & 74.2 \\
\texttt{14b} & 1.1 & y & EE & 2241 & 67 & 60 & 0 & 1 & 0 & 65.4 \\
\texttt{14c} & 1.1 & y & EE & 2244 & 67 & 63 & 0 & 3 & 0 & 65.0 \\
\hline
\texttt{15a} & 1.2 & y & EE & 2244 & 67 & 62 & 0 & 0 & 0 & 40.5 \\
\texttt{15b} & 1.2 & y & EE & 2248 & 53 & 57 & 0 & 0 & 0 & 40.5 \\
\texttt{15c} & 1.2 & y & EE & 2251 & 44 & 67 & 0 & 0 & 0 & 40.5 \\
\hline
\texttt{16a} & 0.8 & n & EE & 2235 & 344 & 25 & 0 & 0 & 0 & 40.5 \\
\texttt{16b} & 0.8 & n & EE & 2244 & 391 & 20 & 0 & 0 & 0 & 40.5 \\
\texttt{16c} & 0.8 & n & EE & 2239 & 344 & 22 & 0 & 0 & 0 & 40.5 \\
\hline
\texttt{17a} & 0.9 & n & EE & 2247 & 290 & 34 & 0 & 0 & 0 & 40.5 \\
\texttt{17b} & 0.9 & n & EE & 2244 & 312 & 23 & 0 & 0 & 0 & 40.5 \\
\texttt{17c} & 0.9 & n & EE & 2249 & 295 & 31 & 0 & 0 & 0 & 40.5 \\
\hline
\texttt{18a} & 1 & n & EE & 2250 & 228 & 33 & 0 & 0 & 0 & 40.5 \\
\texttt{18b} & 1 & n & EE & 2246 & 220 & 38 & 0 & 0 & 0 & 40.5 \\
\texttt{18c} & 1 & n & EE & 2251 & 236 & 28 & 0 & 0 & 0 & 40.5 \\
\hline
\texttt{19a} & 1.1 & n & EE & 2253 & 170 & 48 & 0 & 0 & 0 & 40.5 \\
\texttt{19b} & 1.1 & n & EE & 2252 & 182 & 44 & 0 & 0 & 0 & 40.5 \\
\texttt{19c} & 1.1 & n & EE & 2252 & 160 & 53 & 0 & 0 & 0 & 40.5 \\
\hline
\texttt{20a} & 1.2 & n & EE & 2253 & 145 & 49 & 0 & 0 & 0 & 40.5 \\
\texttt{20b} & 1.2 & n & EE & 2250 & 138 & 49 & 0 & 0 & 0 & 40.5 \\
\texttt{20c} & 1.2 & n & EE & 2248 & 136 & 47 & 0 & 0 & 0 & 40.5 \\
\enddata
\tablecomments{Complete list of all cluster simulations run for this study. In column 2, we list the initial $r_v$ for each simulation. In column 3, we indicate whether or not primordial mass segregation is assumed. In column 4, we indicate the assumed prescription for giant collisions, where ``SS'' indicates the sticky sphere approximation and ``EE'' indicates the envelope ejection prescription, as described in the text. Column 5 shows the total number of BHs retained at birth in each simulation. Column 6 shows the number of these BHs that underwent at least one stellar collision prior to formation. Columns 7, 8, 9, and 10 show the total number of PPSNe, PISNe, mass-gap BHs, and IMBHs in each simulation, respectively. The final column (11) shows the mass of the most massive BH formed in each simulation.}
\end{deluxetable*}

\section{Results}
\label{sec:results}

In this section we describe the results in the context of the formation of BHs in the upper mass gap. In Section \ref{sec:formation} we describe the typical formation pathways to such objects and in Section \ref{sec:population} we decribe the overall features of our complete set of simulations.

\begin{figure}
\begin{center}
\includegraphics[width=0.9\linewidth]{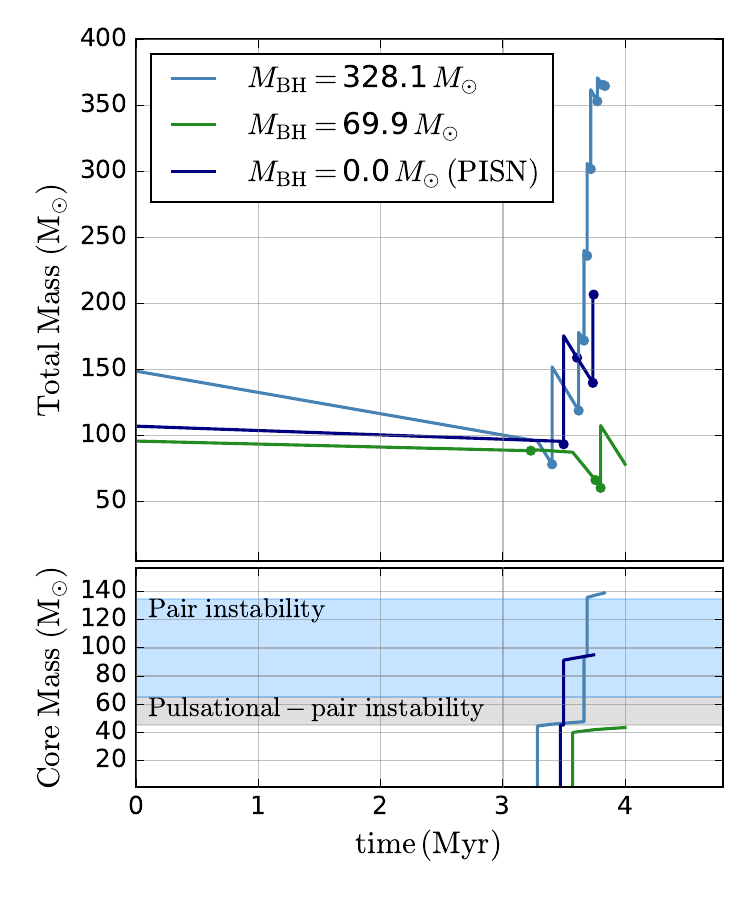}
\caption{\footnotesize \label{fig:mass_vs_time} Total mass (top panel) and core mass (bottom panel) versus time for three collision outcomes in simulation \texttt{2c}. The filled circles in the top panel indicate collision events. The detailed collision histories for each of these three outcomes are shown in Figure \ref{fig:cartoon} and also listed in Tables \ref{table:collision_history}-\ref{table:collision_history3} in the Appendix.}
\end{center}
\end{figure}

\begin{figure*}
\begin{center}
\includegraphics[width=0.8\linewidth]{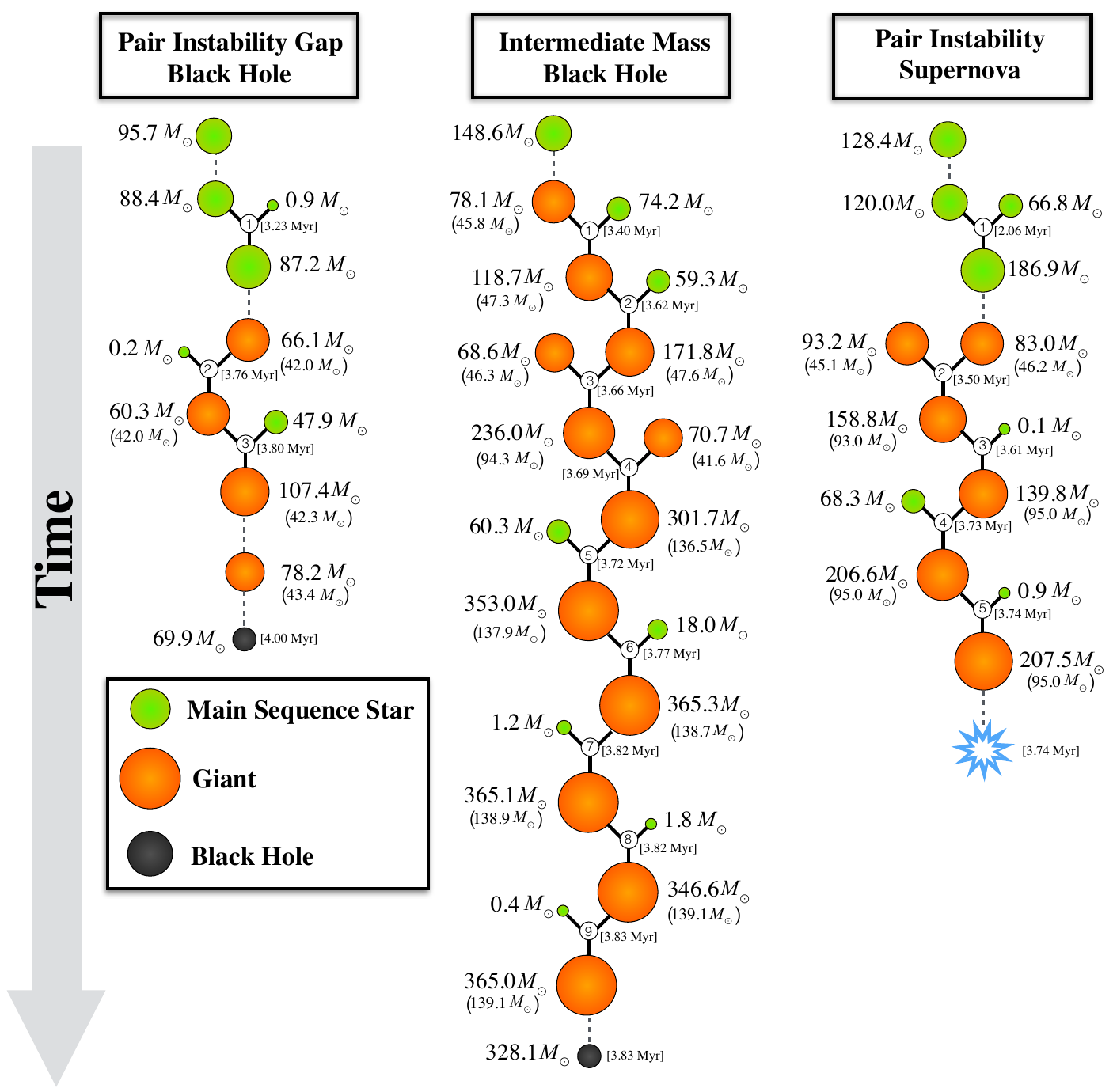}
\caption{\footnotesize \label{fig:cartoon} Example collision histories for the three distinct massive star collision outcomes described in the text. We show the total mass of each object next to each collision component and product. We show in parentheses the core mass of each object and in brackets the time of each collision. The ``Pair Instability Gap Black Hole'', ``Intermediate Mass Black Hole'', and ``Pair Instability Supernova'' histories corresponds to the green, light blue, and navy curves, respectively, in Figure \ref{fig:mass_vs_time}.}
\end{center}
\end{figure*}

\subsection{Evolutionary outcomes from massive stellar collisions}
\label{sec:formation}

\begin{figure*}
\begin{center}
\includegraphics[width=0.95\linewidth]{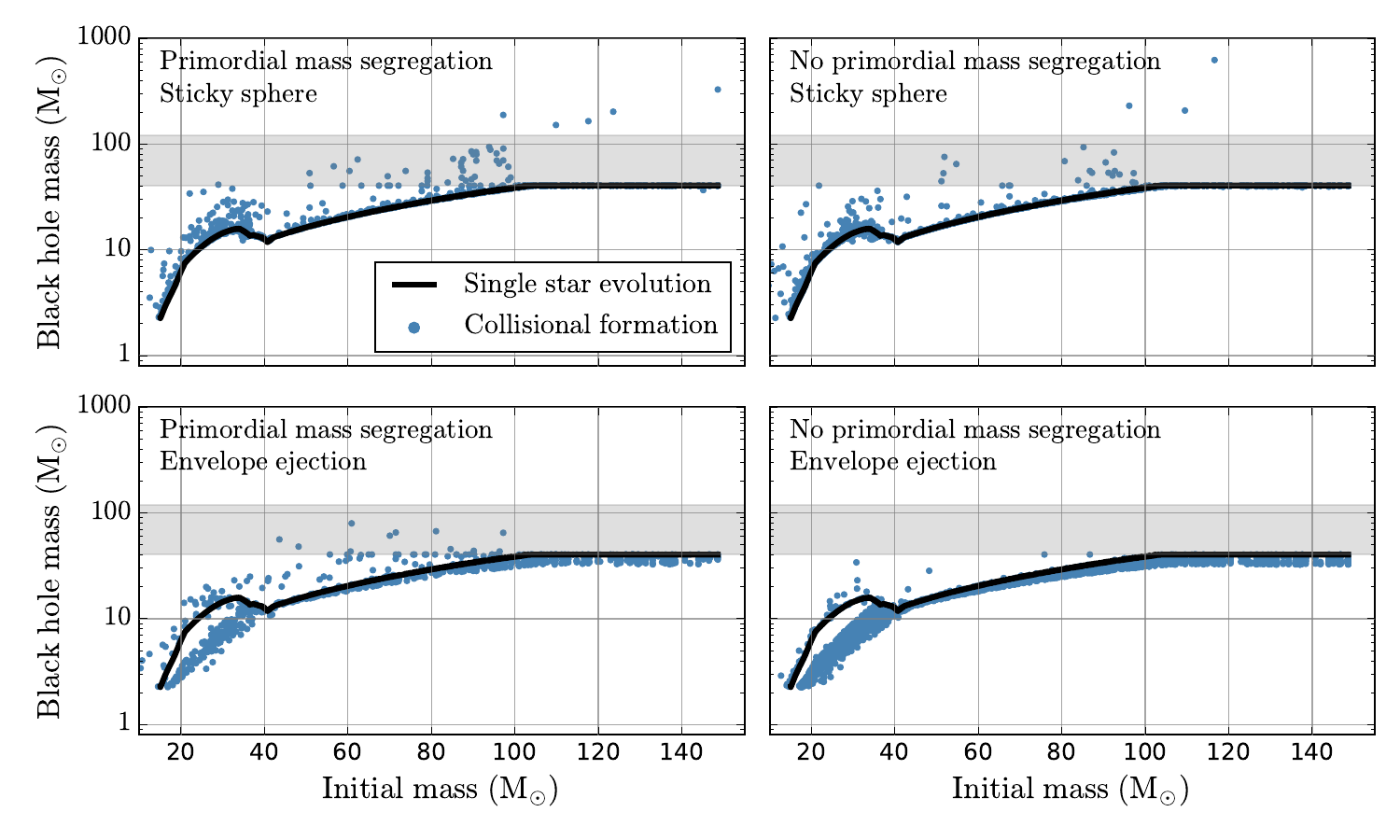}
\caption{\footnotesize \label{fig:BH_vs_ZAMS} BH mass versus initial mass for all BHs formed through stellar collisions (blue points) in the four groups of models. The black curves show the $M_{\rm{BH}}-M_{\rm{ZAMS}}$ tracks for single star evolution and the shaded gray regions indicate the mass gap expected from PISNe. The top-left (top-right) panel shows models which assume primordial mass segregation (no primordial mass segregation) and sticky sphere limit for giant collisions. The bottom-left (bottom-right) panel shows models which assume primordial mass segregation (no primordial mass segregation) and the envelope ejection limit for giant collisions.}
\end{center}
\end{figure*}  

Here, we describe three distinct evolutionary outcomes for massive stars that undergo one or more collisions before stellar core collapse. Each of these outcomes is uniquely dependent upon stellar collisions and will never occur through single star evolution for the assumed IMF.\footnote{In addition to the three discussed outcomes, collisions may also lead to outcomes degenerate with those occurring through single star evolution (i.e., formation of BHs below the pair-instability gap or BHs that form through PPSNe). We discuss these briefly in Section \ref{sec:population}.}

In Figure \ref{fig:mass_vs_time} we show the stellar evolution for a characteristic example of each of the three outcomes (all taken from simulation \texttt{2c} in Table \ref{table:models}, chosen simply because this simulation produces the full range of outcomes). In the top panel, we show the evolution of the total stellar mass up to the moment of stellar core collapse.
Here, each of the filled circles denotes a collision event. In the bottom panel, we show the core mass versus time. The blue (gray) shaded regions marks the mass ranges where PISNe (PPSNe) are assumed to operate. In Figure \ref{fig:cartoon}, we provide cartoon illustrations of the collision sequence for each of the three outcomes shown in Figure \ref{fig:mass_vs_time}. We provide further detail on each collision event in the tables in the Appendix.

Below, we summarize each of these three collision outcomes:

\vspace{0.25cm}

\textit{1. BH in the pair-instability mass gap:} As described in \citet{Spera2019,DiCarlo2019,DiCarlo2020}, if a massive star on the giant branch (i.e., it has a well-developed helium core) undergoes a collision/merger with a second non-evolved star (i.e., a star on the main sequence), the result may be an evolved star with an oversized hydrogen envelope. In particular, if the core mass of this star remains below the minimum mass for PPSN (here assumed to be $45\,M_{\odot}$) and if the star retains a significant fraction of its recently acquired oversized envelope, the ultimate result may be a BH with a mass occupying the upper mass gap. The green curves in Figure \ref{fig:mass_vs_time} illustrate a typical sequence leading to this outcome. See also the left-hand panel of Figure \ref{fig:cartoon}. In total, this outcome occurs 3 times in simulation \texttt{2c}.
    
\vspace{0.25cm}
    
\textit{2. Pair instability supernova:} The most massive stars drawn from the assumed IMF of our simulations is $150\,M_{\odot}$. If left unperturbed, such a star will develop a pre-explosion core mass of just under $50\,M_{\odot}$, falling within the mass range assumed to be subject to PPSNe for our assumed metallicity \citep{Vink2001,Breivik2020}. Thus, for single star evolution alone, PISNe will never occur for our assumed IMF. However, if while on the giant branch, such a massive star undergoes one or more collisions with other giants with similarly massive cores, then the core of the collision product may grow sufficiently to fall in the range assumed to undergo a PISN. In this case, no remnant is formed. The dark blue curve in Figure \ref{fig:mass_vs_time} illustrates a typical sequence leading to this outcome. See also the right-hand panel of Figure \ref{fig:cartoon}. In total, this outcome occurs 5 times in simulation \texttt{2c}. We discuss PISNe in more detail in Section \ref{sec:PISN}.

\vspace{0.25cm}
    
\textit{3. Direct collapse and IMBH formation:} In the event of multiple collisions, the core may grow sufficiently to exceed the maximum core mass assumed to undergo PISNe in which case we assume a direct collapse results. BHs formed through this channel have masses in excess of $\sim120\,M_{\odot}$ and are generally placed in the class of so-called IMBHs.\footnote{The term IMBH is generally used to refer to the class of BHs of mass $\sim10^2-10^5\,M_{\odot}$ that bridge the divide between stellar-mass BHs ($M \lesssim 50\,M_{\odot}$; i.e., upper-limit associated with PPSNe)
and supermassive BHs ($M \gtrsim 10^5\,M_{\odot}$). In this analysis, we use the term ``pair-instability gap'' or ``upper mass gap'' BH to denote those BHs occupying the gap from $\sim40-120\,M_{\odot}$ expected from PPSNe and use the term ``IMBH'' to denote the specific class of massive BHs that form through direct collapse above the PISNe boundary.} The light blue curve in Figure \ref{fig:mass_vs_time} illustrates the formation of an IMBH. See also the middle panel of Figure \ref{fig:cartoon}.
   
In the event of a collisional runaway,
very massive stars in excess of $\sim1000\,M_{\odot}$ may form yielding similarly massive IMBHs \citep[e.g.,][]{Gurkan2004,Ardi2008,Goswami2012,Giersz2015}. A collisional runaway is generally expected to occur if the core mass segregation timescale of massive stars time is less than the stellar lifetime of the massive stars \citep[$t\sim3-5\,$Myr; e.g,][]{Gurkan2004}. We identify collisional runaways in three of our simulations (\texttt{1a}, \texttt{1b}, and \texttt{1c}; see Table \ref{table:models}). Not suprisingly, these three models have the smallest initial $r_v$ (0.8 pc) and therefore the shortest central relaxation times, ideal for triggering collisional runaways. In simulation \texttt{6d} (also $r_v=0.8\,$pc but not assuming primordial mass segregation as in the former three runs) a $623\,M_{\odot}$ IMBH (roughly $700\,M_{\odot}$ pre-collapse progenitor) forms. This object is indeed analogous to the cases in simulations \texttt{1a}, \texttt{1b}, and \texttt{1c}, but does not meet our (admittedly arbitrary) $1000\,M_{\odot}$ requirement for labelling a collisional runaway. In the Appendix, we list full collision histories for each of the three collisional runaways.

\subsection{Population demographics}
\label{sec:population}

Column 5 of Table \ref{table:models} shows the total number of BHs formed and retained at birth in each simulation and column 6 shows the number of these BHs that were formed through stellar collisions. As shown, we find that as many as $20\%$ of all BHs in a typical cluster may have undergone at least one collision prior to collapse. Thus, stellar collisions may play a significant general role in BH formation in GCs. Columns 7--10 of Table \ref{table:models} list the total number of BHs formed through PPSN (assumed to yield BH masses of exactly $40.5\,M_{\odot}$), the number of PISNe, the total number of BHs with masses in the pair-instability gap ($40.5-120\,M_{\odot}$) formed through stellar collisions (see left-hand panel of Figure \ref{fig:cartoon}), and the number of IMBHs with masses in excess of $120\,M_{\odot}$. In addition, column 11 lists the mass of the largest BH formed in each simulation.

In Figure \ref{fig:BH_vs_ZAMS}, we show the BH mass versus initial mass for each BH formed in our simulations. For BHs formed through stellar evolution alone (i.e., they never undergo a collisions prior to core collapse), the initial mass corresponds to the zero-age main sequence (ZAMS) mass. In this case, the $M_{\rm{BH}}-M_{\rm{ZAMS}}$ relation is well-defined and is shown by the solid black curves. For $M_{\rm{ZAMS}}\lesssim 40\,M_{\odot}$, $M_{\rm{BH}}$ is determined by the assumed fallback prescription \citep[here we assume the ``delayed'' model from][]{Fryer2012}, for $M_{\rm{ZAMS}}$ in the range $\approx40-100\,M_{\odot}$, $M_{\rm{BH}}$ is determined primarily by the assumed wind-mass loss model, and for $M_{\rm{ZAMS}} \gtrsim 100\,M_{\odot}$, $M_{\rm{BH}}$ is determined primarily by the assumed pair instability physics.

In blue, we show those BHs that formed through stellar collisions. For these objects, we simply define the initial mass as the ZAMS mass of the more massive of the two collision components at the moment of the first collision in the BH's history. The horizontal gray bands in each panel illustrate the pair-instability mass gap.

\begin{deluxetable*}{cc|c|c|c|c}
\tabletypesize{\scriptsize}
\tablewidth{0pt}
\tablecaption{BH formation efficiency for different prescriptions \label{table:comparison}}
\tablehead{
	\colhead{Prim. MS} &
	\colhead{Giant coll.} &
	\colhead{$N_{\rm{PI}}/M_{\odot}$} &
	\colhead{$N_{\rm{IMBH}}/M_{\odot}$} &
	\colhead{$M_{\rm{PI}}/M_{\odot}$} &
	\colhead{$M_{\rm{IMBH}}/M_{\odot}$}
}
\startdata
y & SS & $4.2\times10^{-6}$ & $5.3\times10^{-7}$ & $2.5\times10^{-4}$ & $1.1\times10^{-4}$ \\ 
n & SS & $2.0\times10^{-6}$ & $3.4\times10^{-7}$ & $1.2\times10^{-4}$ & $1.2\times10^{-4}$ \\
y & EE & $2.8\times10^{-6}$ & $0$ & $1.7\times10^{-4}$ & $0$ \\
n & EE & $0$ & $0$ & $0$ & $0$ \\
\hline
\multicolumn{2}{c|}{Di Carlo et al. 2020} & $4.2\times10^{-5}$ & $2.4\times10^{-6}$ & $3.4\times10^{-3}$ & $5.5\times10^{-4}$\\
\enddata
\tablecomments{In rows 1--4 we list the BH formation efficiency for the various prescriptions adopted in this study. In row 5 we list for comparison the formation efficiency from the lower-mass cluster simulations computed in \citet{DiCarlo2020}. In rows 3 and 4 we list the total number of pair-instability gap BHs ($N_{\rm{PI}}$) and IMBHs ($N_{\rm{IMBH}}$) per unit stellar mass, respectively, and in rows 5 and 6 we list the total mass of pair-instability gap BHs and IMBHs per unit stellar mass, respectively.
}
\end{deluxetable*}

In Table \ref{table:comparison}, we compare BH formation efficiencies (the number of BHs formed per unit stellar mass) for the four different physics prescriptions adopted in our simulations. In columns 3 and 4, we show efficiencies for BHs in the pair-instability gap and IMBHs, respectively. In columns 5 and 6, we show the fraction of the total mass that forms pair-instability BHs and IMBHs, respectively. In the bottom row of Table \ref{table:comparison}, we show for comparison the formation efficiencies computed from the simulations described in \cite{DiCarlo2019} and \cite{DiCarlo2020}.  We discuss comparisons with these earlier analyses in detail in Section \ref{sec:comparison}.

As Figure \ref{fig:BH_vs_ZAMS} and Table \ref{table:comparison} show, BHs with masses within or above the pair-instability mass gap form most readily in the simulations with primordial mass segregation and which treat giant collisions in the sticky sphere approximation. This is as anticipated: primordial mass segregation leads to more massive star collisions and the sticky sphere approximation leads to more significant mass growth during the collisions.

As the lower two panels of Figure \ref{fig:BH_vs_ZAMS} show, we find that the envelope ejection prescription for giant collisions leads to a population of BHs with masses lower that those predicted from single star evolution. In this limit, if a giant star (en route to stellar core collapse and BH formation) collides with another star and loses its envelope, core growth may be inhibited such that at the time of core collapse, the core mass is lower than if the star had evolved uninterrupted. In this case, a lower mass BH results in a process we label as an ``inverse runaway.'' This effect is most pronounced if the giant collides with a low-mass main sequence star, in which case the new giant formed through the collision has an envelope significantly less massive than its pre-collision progenitor.
As pointed out in \citet{Kremer2020}, for a non-mass-segregated cluster where stars of all masses are equally mixed, massive stars are most likely to undergo collisions with low-mass MS stars, simply because these stars dominate the assumed IMF.
However, for a mass-segregated cluster where stars tend to interact with other stars of similar total mass (see Figure \ref{fig:avg_mass}), collisions with mass ratios near unity are more common. 
Thus, in a primordially segregated cluster, giants typically collide with MS stars that are more massive compared to those in non-segregated clusters. As a result, MS--giant collisions preferentially create collision products with a higher envelope mass (therefore yielding more massive BHs) in initially segregated clusters compared to clusters that are not primordially segregated (which produce lower-mass BHs; compare the bottom-left and bottom-right panels in Figure \ref{fig:BH_vs_ZAMS}). Note that even in the ``envelope ejection'' limit, a population of overmassive BHs may still form through the standard runaway process, depending upon the collision histories. Indeed, in the bottom two panels of Figure \ref{fig:BH_vs_ZAMS}, two distinct BH populations are visible, particularly notable for initial masses in the range $20-40\,M_{\odot}$. Here, the location of a given BH in the overmassive versus the undermassive population is determined by its specific collision history.

\section{Implications for Gravitational Wave Astrophysics}
\label{sec:GW}

If the BHs within or above the pair-instability mass gap go onto to merge with one another or with other lower mass BHs through dynamical interactions, these mergers have important implications to GW science. In particular, if such mass-gap mergers are detected by instruments such as LIGO/Virgo, it would provide constraints on the contribution of dynamical environments to the overall binary BH (BBH) merger rate. Indeed, the probability that at least one of the components of the recently detected event GW190521 (inferred component masses of $85^{+21}_{-14}\,M_{\odot}$ and $66^{+17}_{-18}\,M_{\odot}$) falls within the pair-instability mass gap is $99\%$ \citep{LIGO2020a,LIGO2020b}.

In order to investigate this topic, we run seven of the simulations show in Table \ref{table:models} for $12\,$Gyr, recording all BBH mergers that occur. For these simulations, we adopt exclusively the sticky sphere assumption for giant collisions, as this approximation was shown in Section \ref{sec:results} to yield the highest formation rate of mass-gap BHs. In this case, one could regard the results of this section as an upper limit on the true number of collisional mass-gap mergers.\footnote{Again noting the caveat that we assume here zero primordial binaries. Higher binary fractions may increase both the number of pair-instability gap BHs \citep[][see also Section \ref{sec:conclusion}]{DiCarlo2019,DiCarlo2020} and also the number of BBH mergers \citep[e.g.,][]{Chatterjee2017a}.}

\begin{figure}
\begin{center}
\includegraphics[width=\linewidth]{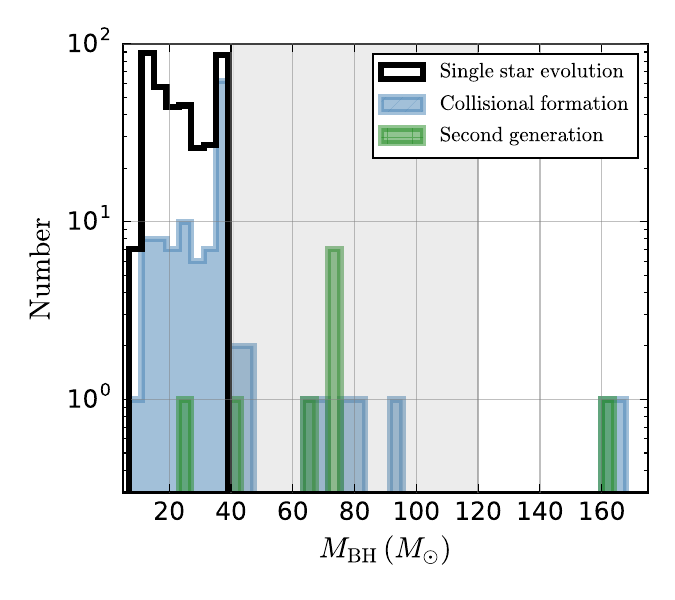}
\caption{\footnotesize \label{fig:BBHmergers} Mass distribution for all BHs that undergo mergers in the seven simulations evolved for $12\,$Gyr. The black histogram shows the merging BHs that form through single star evolution, the blue histogram shows those that form through collisions (see Section \ref{sec:results}), and the green histogram shows the second-generation BHs that formed through earlier BBH mergers. The shaded gray region indicates the mass gap expected from PPSNe and PISNe for single star evolution.}
\end{center}
\end{figure}

In Figure \ref{fig:BBHmergers}, we show the mass distribution for all BHs that undergo BBH mergers in these seven simulations. The black histogram shows those merging BHs that form through single star evolution, the blue histogram shows the BHs that form through collisions of young massive stars (as in Section \ref{sec:results}), and the green histogram shows the second-generation BHs that are formed through earlier BBH mergers. The gray background illustrates the pair-instability mass gap.

Of the 259 total BBH mergers occurring in this set of simulations, 95 ($37\%$) feature at least one component which underwent at least one stellar collision before BH formation. If the stellar collision process encodes itself upon the BH that ultimately forms (for example, by altering the BH's mass, as discussed next, or spin, as discussed briefly in Section \ref{sec:conclusion}), this may yield an observable fingerprint upon potential GW detections.

In total, 16 of the 259 total BBH mergers ($6\%$) involve at least one component with a mass in the pair-instability gap. Of these 16 mass-gap mergers, 7 ($3\%$ of the 259 total) feature at least one BH formed through stellar collisions, and 11 ($4\%$ of the 259 total) feature at least one second-generation BH. Thus, the stellar-collision channel may rival the multiple generation channel as a mechanism for producing mass-gap BBH mergers. We do note that \citet{Rodriguez2019a} identified a larger fraction of second-generation mergers (roughly $10\%$ of all BBH mergers) compared to the $4\%$ identified here. This is not surprising given that \citet{Rodriguez2019a} examined more massive clusters models (up to $\sim10^6\,M_{\odot}$ at birth, over twice as massive as the models considered in this study). Due to their higher escape velocities, more massive clusters retain a larger fraction second-generation BHs upon formation. We can speculate that more massive (and therefore more dense) clusters will also yield a larger number of massive star collisions, and therefore more collisional mass-gap BHs, but more detailed models are necessary to test this. We discuss this topic further in Section \ref{sec:conclusion}.

In Table \ref{table:merger_list}, we list the properties of these 16 mass-gap mergers, including merger time (relative to the host cluster birth time), component masses, and merger channel. In this table, BH masses marked with a $\star$ are formed through stellar collisions while those marked with a $\dagger$ are formed through BBH mergers. In column 5, we show the channel through which each listed BBH merger occurs. Here ``binary'' mergers are in-cluster mergers occurring through two-body GW inspiral, ``binary--single'' and ``binary--binary'' mergers are in-cluster mergers that occur through close passages during 3- and 4-body resonant encounters, and ``ejected'' mergers are those that are dynamically ejected from their host clusters and merge as isolated binaries. See \citet{Samsing2018a,D'Orazio2018,Rodriguez2018b,Zevin2018,Kremer2020} for a summary of the different BBH merger channels in dense star clusters.

In column 6 in Table \ref{table:merger_list}, we note whether each merger is ejected or retained post merger (determined by the relative value of the computed GW recoil kick and current cluster escape velocity). Mergers marked ``N/A'' were ejected from their host cluster prior to merger. Of the 7 BBH mergers listed that occur in their host cluster, all but one are ejected post merger through GW recoil. Indeed, the one mass-gap merger product that is retained (the $94\,M_{\odot}+80\,M_{\odot}$ merger where both components are formed through stellar collisions) is ejected from the cluster shortly thereafter through a binary--binary resonant encounter. Thus, as has been shown through various other analyses \citep[e.g.,][]{Antonini2019}, GW recoil kicks combined with the relatively low escape velocities of typical GCs prevents the build up of massive BHs (i.e., IMBHs) through repeated BH mergers. As discussed in Section \ref{sec:results}, runaway stellar mergers is a far more viable channel for producing IMBHs beyond with masses far in excess of the pair-instability mass gap.

Previous analyses \citep[e.g.,][]{Ziosi2014,Gnedin2014,RodriguezLoeb2018,Fragione2018b,Banerjee2018,DiCarlo2019,Kremer2020} have predicted the overall BBH merger rate from clusters as high as $20\,\rm{Gpc}^{-3}\,\rm{yr}^{-1}$ in the local universe, within the uncertainty bounds of the merger rates inferred from the LIGO/Virgo O2 catalog, $53.2^{+58.5}_{-28.8}\,\rm{Gpc}^{-3}\rm{yr}^{-1}$ \citep{LIGO2018a,LIGO2018b}. We find that up to roughly $37\%$ of all BBH mergers in clusters may host at least one BH formed through stellar collisions and that up to roughly $3\%$ of all mergers include at least one mass-gap BH that formed through stellar collisions. From these numbers, we can infer a rough estimate of the rate of mass-gap mergers of $0.6\,\rm{Gpc}^{-3}\rm{yr}^{-1}$ from stellar collisions in clusters. We note this is consistent with the $\approx 0.1\,\rm{Gpc}^{-3}\rm{yr}^{-1}$ rate inferred by LIGO/Virgo for GW190521-like events, but caution that more detailed analysis is necessary to make precise comparisons and rate predictions. Regardless, we can conclude that the overall rate of collisional BH mergers may constitute an observable fraction of GW events detected by LIGO/Virgo.

\begin{deluxetable}{c|c|c|c|l|l}
\tabletypesize{\scriptsize}
\tablewidth{0pt}
\tablecaption{All binary BH mergers with at least one component in upper mass gap \label{table:merger_list}}
\tablehead{
	\colhead{Model} &
	\colhead{$t_{\rm{merge}}$} &
	\colhead{$M_1$} &
	\colhead{$M_2$} &
	\colhead{type} &
	\colhead{outcome}\\
	\colhead{} &
	\colhead{(Gyr)} &
	\multicolumn{2}{c}{$M_{\odot}$} &
	\colhead{} &
	\colhead{}
}
\startdata
2a & 0.06 & $94.2^{\star}$ & $80.1^{\star}$ & binary & Retained\\
2a & 0.15 & $71.6^{\star}$ & $77.1^{\dagger}$ & binary-binary & Ejected\\
2a & 0.21 & 36.4 & $47.6^{\star}$ & binary-single & Ejected\\
2a & 1.37 & 25.4 & $53.3^{\dagger}$ & binary & Ejected\\
2a & 1.7 & 40.5 & $165.9^{\star,\dagger}$ & ejected & N/A\\
\hline
2b & 0.71 & 40.5 & $168.0^{\star}$ & ejected & N/A\\
\hline
3a & 0.15 & $66.1^{\star}$ & $76.2^{\dagger}$ & binary & Ejected\\
3a & 0.21 & 33.3 & $77.1^{\dagger}$ & binary & Ejected\\
3a & 12.92 & 33.9 & $73.8^{\dagger}$ & ejected & N/A\\
\hline
4a & 0.55 & 40.5 & $77.1^{\dagger}$ & ejected & N/A\\
\hline
5a & 0.46 & 38.1 & $45.4^{\star}$ & ejected & N/A\\
5a & 7.77 & 33.9 & $76.8^{\dagger}$ & ejected & N/A\\
\hline
5b & 0.68 & 33.9 & $74.2^{\dagger}$ & binary-binary & Ejected\\
5b & 9.8 & 40.5 & $85.0^{\star}$ & ejected & N/A\\
\hline
6a & 0.29 & 40.5 & $73.9^{\dagger}$ & ejected & N/A\\
6a & 11.61 & 36.3 & $41.5^{\dagger}$ & ejected & N/A\\
\hline
\enddata
\tablecomments{List of all BBH mergers in the seven simulations integrated to $12\,$Gyr with at least one component in the pair-instability mass gap. BH masses with marked with a $\star$ are formed through stellar collisions and BH masses marked with a $\dagger$ are second-generation BHs formed through previous BBH mergers. The $165.9\,M_{\odot}$ object marked with both symbols was formed through the merger of a pair of BHs one of which was itself formed through collisions. The fifth column notes the merger channel for each binary (see text for details) and the sixth column denotes the merger outcome (if the merger product is retained post-merger or is ejected due to the GW recoil kick).}
\end{deluxetable}

\section{Pair-instability supernovae and other electromagnetic transients}
\label{sec:PISN}

In this section, we compute rates (Section \ref{sec:rates}) and discuss observational prospects (Section \ref{sec:obs_prospects}) for PISNe and a number of other possible electromagnetic transients associated with the outcomes of stellar collisions seen in our simulations. Because we identify PISNe only in those simulations adopting the sticky sphere limit for giant collisions, we consider only these simulations in this section (simulations \texttt{2-10} in Table \ref{table:models}; neglecting simulations \texttt{1a-c} that undergo collisional runaways). In this case, the rate predictions presented in this section may be regarded as upper limits.

\subsection{Rates}
\label{sec:rates}

In order to estimate the cosmological rates of various SN events, we adopt a method similar to that implemented in \citet{Kremer2020} to compute BBH merger rates. Here the cumulative rate is given by:

\begin{equation}
    R(z) = \int_0^z \mathcal{R}(z^\prime) \frac{dV_c}{dz^\prime}(1+z^\prime)^{-1}dz^\prime,
\end{equation}
where $dV_c/dz$ is the comoving volume at redshift $z$ and $\mathcal{R}(z)$ is the comoving (source) rate given by $\mathcal{R}(z) = \rho_{\rm{GC}} \times \frac{dN(z)}{dt}$, where $\rho_{\rm{GC}}$ is the volumetric number density of clusters, assuming a constant value of $\rho_{\rm{GC}}=2.31\,\rm{Mpc}^{-3}$ \citep[consistent with][]{Rodriguez2015a,RodriguezLoeb2018,Kremer2020} and $dN(z)/dt$ is the number of events per unit time at a given redshift.

We compute $dN(z)/dt$ using a procedure similar to that of \citet{Kremer2020}: first, we generate a complete list of event times ($t_{\rm{SN}}$) for all SNe occurring in our model set. For each of these events, we draw 200 random ages ($t_{\rm{age}}$) for the host cluster in which the SN occurred. We then compute the effective event time for each SN as $t_{\rm{effective}} = t_{\rm{Hubble}} - t_{\rm{age}} + t_{\rm{SN}}$. We draw cluster ages from the age distributions of \citet{El-Badry2018}.\footnote{Note that cluster age distributions are metallicity-dependent \citep[see][]{El-Badry2018}, however the simulations of this study adopt a fixed metallicity. For simplicity, we ignore metallicity effects for the rate calculation, and refer the reader to \citet{DiCarlo2020} for a discussion of the effect of metallicity in regards to stellar collision outcomes in dense clusters.} We then compute the number of events per time, $dN(z)/dt$, by dividing this list of effective event times into separate redshift bins, accounting for the oversampling of age draws and total number of simulations (35 total that adopt the sticky sphere limit; see Table \ref{table:models}).

The cluster simulations in this analysis have initial masses of roughly $4\times10^5\,M_{\odot}$ and present-day (i.e., at $t\approx12\,$Gyr) of roughly $2\times10^5\,M_{\odot}$\footnote{The cluster mass loss is governed by both stellar wind mass loss and dynamical ejection of stars throughout the cluster's lifetime.}, matching well the median cluster mass observed for the Milky Way GCs \citep[e.g.,][]{Harris1996,Baumgardt2018,Kremer2020}. However, as described in \citet{Rodriguez2015a,Kremer2020} in the case of BBH mergers, adopting this simulation mass as our typical cluster causes an underestimate of the total rate, because it does not properly account for the contribution of the cluster mass function's high-mass tail not covered by our models, which as shown in \citet{Kremer2020} may yield a rate higher by a factor of a few. A careful examination of the way the total number of PISNe and other SN events per model scales with $N$ is beyond the scope of this paper. We simply note that the rates presented here may in factor underestimate the true rate of these events.

\begin{figure}
\begin{center}
\includegraphics[width=0.9\linewidth]{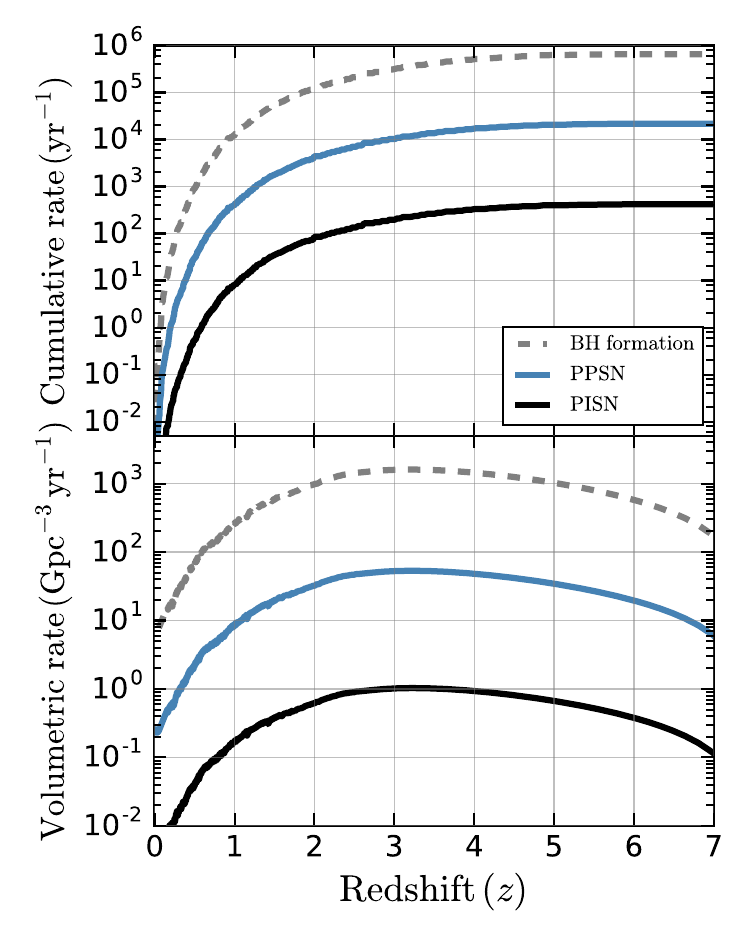}
\caption{\footnotesize \label{fig:rates} Cumulative (top panel) and volumetric (bottom panel) rates for various classes of SNe as computed from our simulations.}
\end{center}
\end{figure}

We show in Figure \ref{fig:rates} the results of these rate calculations for PISNe (black curve) and for PPSNe (blue curve). We also show (dashed gray cure) the formation rate of all BHs, which may be associated with standard core-collapse SNe.
Note that because BH formation occurs exclusively at early times ($t\approx3-20\,$Myr), the shape of these curves are roughly identical and are determined primarily by the assumed cluster birth time distribution.

\subsection{Observational prospects}
\label{sec:obs_prospects}

As described in Section \ref{sec:method}, PISNe are thought to occur for stars with helium core masses in the range $\approx65-135\,M_{\odot}$. At zero metallicity (i.e., Pop III stars), this corresponds to ZAMS masses in the range $\sim140-260\,M_{\odot}$ \citep[e.g.,][]{Spera2017}. At higher metallicities where line-driven stellar winds \citep[e.g.,][]{Vink2001} are expected to drive efficient mass loss, even larger ZAMS masses are likely required to produce helium cores within the range required for a PISN. However, myriad observations suggest an upper limit to the stellar IMF for Pop I/II stars of roughly $150\,M_{\odot}$ \citep[e.g.,][]{Kroupa2001,Figer2005,Weidner2010,Bastian2010}. This may suggest that PISNe are prevented for Pop I/II and thus prevented at redshifts $z \lesssim 6$ which corresponds roughly to the end of the reionization epoch at which the universe become sufficiently metal-enriched to prohibit Pop III star formation \citep[e.g.,][]{BrommLoeb2003}.

From an observational perspective, PISNe are expected to be characterized by very bright (as high as $\sim10^{44}\,\rm{erg\,s}^{-1}$) and broad lightcurves (rise times as long as $\sim150\,$days) as a consequence of the large ejecta masses and, therefore, long diffusion times \citep[e.g.,][]{Dessart2013,Kozyreva2014}. A number of observed events have been associated with PISNe, including SN 2006gy \citep{Smith2007}, SN2007bi \citep{Gal-Yam2009}, SN 2213-1745 \citep{Cooke2012}, OGLE14-073 \citep{Terreran2017}, SN 2016iet \citep{Gomez2019}, and SN2016aps \citep{Nicholl2020}. Several of these candidates have been observed at low redshift, implying that potential formation pathways for PISNe do indeed exist for higher-metallicity Pop I/II stars.

As demonstrated here and in other earlier analyses \citep[e.g.,][]{Langer2007,PortegiesZwart2007,Yungelson2008,Glebbeek2009,Pan2012} the formation of massive stars through stellar collisions in dense star clusters may provide a viable pathway to PISNe. In total, we predict a volumetric rate of roughly $0.1 \rm{Gpc}^{-3}\rm{yr}^{-1}$ in the local universe ($z<1$), roughly consistent with the rate predicted in \citet{Pan2012} which adopted a simple analytic rate estimate using theoretical estimates of cluster and stellar properties. 
We reserve a more detailed examination of the electromagnetic signatures (e.g., lightcurves) of the PISNe (as well as PPSNe and other SN types) identified in our simulations for a later study. We do note however, that \citet{Pan2012} estimated, using simulated PISN lightcurves from \citet{Kasen2011}, that LSST may observe $\sim100$ PISNe per year that originated from collisional runaways in young massive clusters. Thus, in the coming years LSST may indeed place further constraints upon the processes discussed here. 

\section{Discussion and conclusions}
\label{sec:conclusion}

\subsection{Summary}

We have explored ways in which stellar collisions (or series of stellar collisions) in dense star clusters  may lead to several unique evolutionary outcomes not possible for single star evolution. In particular, we have examined the role played by stellar collisions in navigating the gap expected in the BH mass function caused by the pair instability and pulsational pair instability. By computing a large set of independent cluster simulations with cluster masses ($M= 4\times 10^5\,M_{\odot}$) comparable to the GCs observed in the Milky Way, we have demonstrated the dynamical formation of (1) BHs with masses occupying the mass gap expected due to PPSNe and PISNe, (2) massive stars that undergo PISNe, and (3) massive stars with helium cores beyond the boundary where the pair-instability is expected to operate which directly collapse into IMBHs with masses in excess of $\sim100\,M_{\odot}$.

We explored also the dependence of these three outcomes upon two theoretical uncertainties: the degree of primordial mass segregation in the host clusters and the efficiency of envelope ejection during giant star collisions. Together, these two theoretical uncertainties roughly bracket the range of expected outcomes. On the one extreme (assuming complete primordial mass segregation and full sticky sphere collisions), massive BHs readily form through successive stellar collisions. On the other hand (assuming no primordial mass segregation and that giant envelopes are ejected), we observe ``inverse runaways'' where successive collisions can lead to stripping which ultimately may produce lower-mass BHs.

We showed that the population of BHs with masses in or above the pair-instability gap go onto form binaries and merge with other BHs, creating a unique class of upper-mass-gap BBH mergers that may be detectable as GW sources by LIGO/Virgo, similar to the recent event GW190521.
We showed in particular that the collisional formation scenario studied here may compete with the previously explored multiple-generation-merger channel for producing BHs in the pair-instability gap. 

Finally, we computed the volumetric merger rate of a number of SN classes originating in massive star clusters, most notably PISNe which result uniquely from stellar collisions. A number of observed transients have been speculatively linked with PISNe. In the coming years, LSST may provide further constraints upon the potential role of stellar clusters in producing PISNe and other transients.

\subsection{Comparison with previous results}
\label{sec:comparison}

Prior to our present work in the massive star cluster regime, \cite{DiCarlo2019} and \cite{DiCarlo2020} explored the formation of massive BHs in relatively low-mass young star clusters. In this section, we compare our results to those of these previous studies. Such comparisons provide a critical test of our results, help determine the role of differing physical prescriptions across multiple cluster dynamics codes, and, ultimately, constrain ways massive BH formation efficiency varies across the cluster mass function.

\citet{DiCarlo2019} and \citet{DiCarlo2020} computed a set of $5\times10^3$ direct N-body simulations of young star clusters with metallicity $Z=0.002$ and initial masses ranging from $10^3\msun{}$ to $3\times10^4\msun{}$. These simulations were performed using the code \texttt{NBODY6++GPU} \citep{Wang2015}, coupled with the population synthesis code \texttt{MOBSE} \citep{Giacobbo2018}. There are several differences between \texttt{MOBSE} and the \texttt{BSE} stellar-evolution prescriptions implemented in \texttt{CMC} that are relevant for stellar collisions and massive BH formation. First, the rejuvenation factor $f_{\rm{rejuv}}$ is set equal to $0.1$ in \texttt{MOBSE}, meaning that the collision products are more rejuvenated than in \texttt{CMC}. The mass of the collision product in \texttt{MOBSE} is obtained as in the sticky-sphere prescriptions in \texttt{CMC}, with the exception that if two main sequence stars collide, the mass of the collision product is $M_3=M_1+0.7M_2$. Additionally, \texttt{MOBSE} uses the fitting formulae provided by \cite{Spera2017} to determine the remnant mass after a PPSN. The implementation of such formulae in \texttt{MOBSE} is described in the appendix of \cite{Giacobbo2018}. As a result, the \texttt{MOBSE} pair-instability gap for BHs occupies the range $60-150\,M_{\odot}$, compared to \texttt{CMC} where the assumed range is $40.5-120\,M_{\odot}$.

As shown in Table \ref{table:comparison}, the efficiencies of the Di Carlo et al. analyses are roughly a factor of 10 larger than those predicted from the \texttt{CMC} simulations.  There are two primary reasons for this result. First, unlike the \texttt{CMC} simulations which assume zero primordial binaries, the Di Carlo et al. simulations adopt a primordial binary fraction $f_{\rm{bin}}=0.4$.  As shown in a number of previous analyses \citep[e.g.][]{Fregeau2007}, higher binary fractions lead to a higher rate of stellar collisions simply because binaries have a larger cross section for interaction. Second, the Di Carlo et al. simulations adopt fractal initial conditions in order to mimic the clumpy and asymmetric structure observed in star forming regions \citep[e.g.,][]{CartwrightWhitworth2004, Gutermuth2005}. Coupled with their assumed initial half-mass radii \citep[derived from the Marks \& Kroupa relation][]{Marks2012}, these fractal clumps produce regions with density $\rho>10^6\,\msun/\rm{pc}^3$. As shown in Figure \ref{fig:avg_mass}, this is higher than the densities of our \texttt{CMC} simulations in all but the innermost cluster regions. These overdense fractal regions combined with the higher binary fractions lead to an increased rate in stellar collisons in the Di Carlo et al. simulations, thus leading to a higher formation efficiency of both mass-gap BHs and IMBHs.

Using the results of Section \ref{sec:results} and normalizing by total simulated mass, we find an overall BBH merger efficiency (i.e., number of mergers per stellar mass) of $8\times10^{-5}\,M_{\odot}^{-1}$ for the \texttt{CMC} simulations. For mergers in which at least one component is a pair-instability gap BH formed through collisions (multiple BH mergers), the merger efficiency is $2\times10^{-6}\,M_{\odot}^{-1}$ ($3\times10^{-5}\,M_{\odot}^{-1}$). In contrast, the overall BBH merger efficiency from the Di Carlo et al. simulations is $1.3\times10^{-5}\,M_{\odot}^{-1}$, while the merger efficiency with at least one component in the pair-instability gap is $2.5\times10^{-7}\,M_{\odot}^{-1}$. Thus, although the Di Carlo et al. simulations are more efficient (by a factor of $\sim10$) in producing pair-instability gap BHs, they are less efficient (by a factor of $\sim10$) in producing pair-instability gap mergers. This is anticipated. In the Di Carlo et al. simulations (which have lower $N$ compared to the \texttt{CMC} models), a significant amount of the available mass of high-mass stars must be utilized to create a single massive BH. Thus, in the case that a massive BH is formed, the number of available companions in the host cluster is fewer. However, in the \texttt{CMC} simulations, there are many available companions both because $N$ is larger and because the escape velocity is larger, allowing the cluster to retain a larger fraction of lower-mass BHs that may be kicked out of lower-mass clusters through natal kicks.

As can be read from the fifth column of Table \ref{table:merger_list}, 7 of the 16 total mass-gap mergers seem in the \texttt{CMC} simulations occur within their host cluster. This in contrast to the Di Carlo et al. simulations where \textit{all} BBH mergers occur after ejection from their host cluster (including also most of the low-mass mergers where both components lie below the pair-instability gap). This is simply because the low-mass Di Carlo et al. simulations have lower escape velocities. As a consequence, the Di Carlo et al. simulations produce no second-generation BHs. This is a key difference between low-mass clusters and the higher-mass clusters considered in this study.

Ultimately, a more complete study implementing self-consistent binary evolution, stellar collision prescriptions, and cluster initial conditions is needed to determine more precisely the differences between the low-mass simulations of \citet{DiCarlo2019,DiCarlo2020} and the high-mass clusters of the present study.

\subsection{Future work}

A number of elements have been left unexplored in detail in this analysis. Here, we briefly summarize a few such points and describe several avenues for future study. 

We have assumed a fixed metallicity of $Z=0.002 = 0.1Z_{\odot}$ for all simulations computed in this study. Previous studies \citep[e.g.,][]{Glebbeek2009,DiCarlo2020} have shown that metallicity can have a significant effect upon growth through stellar collisions, and therefore upon the mass of the BH ultimately formed. Specifically, both of the aforementioned analyses showed that at higher metallicities (especially approaching $Z_{\odot}$) mass growth can be significantly limited. In the context of low mass clusters, \citet{DiCarlo2020} showed in particular that at solar metallicity, roughly an order-of-magnitude fewer pair-instability gap BHs form compared to simulations with $0.1Z_{\odot}$. This is driven primarily by the assumed metallicity-dependent wind mass loss prescriptions \citep{Vink2001}. 
Furthermore, aside from metallicity dependencies, stellar winds may operate very differently for massive stars ($M \gtrsim 150\,M_{\odot}$) compared to lower mass stars. An alternative treatment of massive star winds may have a substantial effect on the evolution of collision products, particularly for collisions of main-sequence stars \citep[e.g.,][]{Glebbeek2009,Chatterjee2009}. 
We reserve a more detailed examination of the effects of stellar winds on massive BH formation for future studies.

In the simulations computed in this study, we assume for simplicity that BHs are born with zero natal spin (dimensionless spin parameter $a=0$). This is in part motivated by recent work \citep{FullerMa2019} suggesting that stellar-mass BHs are born with low spins, however other work \citep[for example analysis of the spins of BHs found in high-mass X-ray binaries;][]{MillerMiller2015,Fragos2015} suggest some BHs may in fact be born with high spins. In reality, the true values of BH spins remain highly uncertain \citep[e.g.,][]{Heger2005,LovegroveWoosley2013,Qin2019}. However, it is understood that the GW recoil of a BBH merger product is highly sensitive to the spins (both spin magnitude and relative orientation) of the merger components \citep[e.g.,][]{Merritt2004,Campanelli2007,Berti2007,Lousto2012,GerosaBerti2019}. In general, as spin magnitudes increase, the recoil velocity increases. For in-cluster BBH mergers, this means that rapidly spinning BHs are more likely to be ejected from their host cluster upon merger, thus inhibiting the rate of second (and higher) generation mergers. Indeed, \citet{Rodriguez2019a} showed that even for dimensionless spin parameters of $a=0.2$, the rate of second-generation BBH mergers with at least one component in the pair-instability gap nearly vanishes. Thus, if in fact some (or all BHs) are born with non-zero spins, the stellar-collision channel may in fact dominate over the second-generation merger channel in terms of the overall rate of mass-gap BBH mergers from dense clusters.

On this note, we have shown that roughly $37\%$ of all BBH mergers feature at least one BH component which was formed though stellar collisions. Although highly uncertain, we can speculate that these collisions may lead to stellar spin-up (if the collision is off-center, some fraction of the orbital angular momentum of the pair of stars may be transferred to spin angular momentum of the collision product). This may affect the spin of the BH when the collision product ultimately undergoes core collapse. Along these lines, \citet{Batta2019} explored the masses and spins of BH remnants formed through the collapse of rotating, helium star pre-SN progenitors and showed that progenitor stars with rotation rates large enough to form an accretion disk may unbind their outer layers through accretion feedback and produce BHs with only a fraction of the total mass of their progenitors. Furthermore, \citet{Gaburov2010} showed that mass loss during stellar collisions may result in stellar kicks of $\sim10\,\rm{km\,s}^{-1}$. Such kicks, which have not been considered here, may affect the long-term dynamics of the collision products and may be particularly important in lower-mass clusters with lower escape speeds. Ultimately, all these speculations should be tested with more detailed hydrodynamic models capable of computing spin angular momenta of collision products \citep[e.g.,][]{Lombardi2002} coupled with more detailed stellar evolution models \citep[e.g., \texttt{MESA};][]{Paxton2015}.

For most of the cluster simulations considered in this study, we identify only 1-4 BHs in the pair-instability gap with large variations in the maximum BH mass, even between multiple realizations of the same initial conditions. The present study aims to simply demonstrate that the formation of pair-instability gap BHs and IMBHs through stellar collisions is, in principle, possible in GCs. Given the small number statistics at play, a more expansive set of models, with a large number of independent realizations per set of initial conditions, may be necessary to pin down more precisely the typical masses and numbers massive BHs.

We have assumed here a standard IMF from \citet{Kroupa2001} for all models, consistent with a large body of previous cluster modelling work. However, some recent analyses suggest that a top-heavy IMF may be appropriate in some contexts \citep[e.g.,][]{Marks2012,Schneider2018}. A top-heavy IMF will lead to increased rate of massive star collisions and thus may lead to an increased rate of massive BH formation (both pair-instability gap BHs and IMBHs). We reserve detailed exploration of the effect of IMF variations for future work.

Finally, in several of the simulations computed in this study, we demonstrated the formation of very massive stars in excess of $1000\,M_{\odot}$ that may ultimately directly collapse into IMBHs of comparable mass. IMBHs have long been a hotly debated topic due to their potential role in not only GC dynamics \citep[e.g.,][]{Greene2019} but also in cosmology and galaxy formation, as they could be the seeds for the supermassive BHs observed at the centers of most galaxies \citep[e.g.,][]{Katz2015}.
In spite of the inherent interest in these objects, observational evidence for the presence of an IMBH in any GC, either from X-ray and radio observations \citep[e.g.,][]{Tremou2018} or from dynamical measurements \citep[e.g.,][]{Feldmeier2013,Lutz2011,Noyola2010, Perera2017} remains controversial \citep[e.g.,][]{Gieles2018,Zocchi2019}. Nonetheless, the role that IMBHs, if present, may play in the production of GW sources \citep[e.g.,][]{Amaro-Seoane2007,Mandel2008,MacLeod2016b,Fragginskoc2018,Fragione2020, MariaHolz2020}, high-energy transients such as tidal disruption events \citep[e.g.,][]{Rosswog2009,MacLeod2014,MacLeod2016b,Fragleiginskoc2018}, and GC dynamics more broadly is a rich topic that we hope to explore in more detail within the scope of \texttt{CMC} in a later study. 

\acknowledgements{This work was supported in part by NSF Grant AST-1716762 at Northwestern University. KK is supported by an NSF Astronomy and Astrophysics Postdoctoral Fellowship under award AST-2001751. MS acknowledges funding from the European Union’s Horizon 2020 research and innovation programme under the Marie-Sk\l{}odowska-Curie grant agreement No. 794393. SC acknowledges support of the Department of Atomic Energy, Government of India, under project no. 12-R\&D-TFR-5.02-0200. GF acknowledges support from a CIERA Fellowship at Northwestern University. CR was supported by an ITC Postdoctoral Fellowship from Harvard University. This work used computing resources at CIERA funded by NSF PHY-1726951.}

\software{\texttt{CMC} \citep{Joshi2000,Joshi2001,Fregeau2003, Fregeau2007, Chatterjee2010,Chatterjee2013,Umbreit2012,Morscher2013,Rodriguez2018b, Kremer2020}, \texttt{Fewbody} \citep{Fregeau2004}, \texttt{COSMIC} \citep{Breivik2020}}

\bibliographystyle{aasjournal}
\bibliography{mybib}

\appendix

We include in the Appendix the complete collision history for several BHs formed in our simulations. Tables \ref{table:collision_history}--\ref{table:collision_history3} show histories for the three outcomes shown in Figures \ref{fig:mass_vs_time} and \ref{fig:cartoon}. Tables \ref{table:collision_history4}--\ref{table:collision_history6} show histories for the collisional runaways that occur in simulations \texttt{1a}, \texttt{1b}, and \texttt{1c}. Note that in columns 3-5 of these tables, star type ($k$) of 0 refers to deeply or fully convective low mass ($M\leq0.7\,M_{\odot}$) main sequence stars, $k=1$ refers to main sequence stars with $M>0.7\,M_{\odot}$, $k=4$ refers to core helium burning stars, and $k=5$ refers to stars on the first asymptotic giant branch \citep[all following the nomenclature of \texttt{SSE};][]{Hurley2000}.

\newpage

\begin{deluxetable}{c|c|ccc|ccc|ccc|cc}
\tabletypesize{\scriptsize}
\tablewidth{0pt}
\tablecaption{Collision history for IMBH; simulation \texttt{2c}.\label{table:collision_history}}
\tablehead{
	\colhead{} &
	\colhead{Time} &
	\colhead{$k_1$} &
	\colhead{$k_2$} &
	\colhead{$k_3$} &
	\colhead{$M_1$} &
	\colhead{$M_2$} &
	\colhead{$M_3$} &
	\colhead{$M_{\rm{core},1}$} &
	\colhead{$M_{\rm{core},2}$} &
	\colhead{$M_{\rm{core},3}$} &	
    \colhead{$b$} &
	\colhead{$v_{\infty}$}\\
	\colhead{} &
	\colhead{(Myr)} &
	\colhead{} &
	\colhead{} &
	\colhead{} &
	\multicolumn{3}{c}{($M_{\odot}$)} &
	\multicolumn{3}{c}{($M_{\odot}$)} &
    \colhead{($R_{\odot}$)} &
	\colhead{($\rm{km\,s}^{-1}$)}
}
\startdata
1 & 3.402 & 1 & 4 & 4 & 74.2 & 78.0 & 151.8 & 0.0 & 45.8 & 45.8 & 16593.1 & 45.0 \\
2 & 3.619 & 1 & 4 & 4 & 59.3 & 119.0 & 177.9 & 0.0 & 47.3 & 47.3 & 16148.1 & 44.9 \\
3 & 3.663 & 4 & 4 & 4 & 68.6 & 172.0 & 240.0 & 46.2 & 47.6 & 93.9 & 13960.6 & 38.7 \\
4 & 3.689 & 4 & 4 & 4 & 70.7 & 236.0 & 305.8 & 41.6 & 94.3 & 135.9 & 139652.0 & 7.8 \\
5 & 3.718 & 1 & 4 & 4 & 60.3 & 302.0 & 361.7 & 0.0 & 136.5 & 136.5 & 70966.5 & 20.7 \\
6 & 3.772 & 1 & 4 & 4 & 18.0 & 353.0 & 370.8 & 0.0 & 137.9 & 137.9 & 12204.0 & 31.8 \\
7 & 3.816 & 1 & 4 & 4 & 1.2 & 365.0 & 366.5 & 0.0 & 138.7 & 138.7 & 22870.3 & 53.3 \\
8 & 3.823 & 1 & 4 & 4 & 1.8 & 365.0 & 366.9 & 0.0 & 138.9 & 138.9 & 12629.0 & 77.2 \\
9 & 3.834 & 0 & 5 & 5 & 0.4 & 365.0 & 365.0 & 0.0 & 139.1 & 139.1 & 50813.6 & 59.1 \\
\hline
10 & 3.834 & - & - & 14 & - & - & 328.1 & & & & &  \\
\enddata
\tablecomments{Collision history for the $328.1\,M_{\odot}$ IMBH shown in Figure \ref{fig:mass_vs_time}. The numbers in column 1 correspond to the numbered events shown in Figure \ref{fig:cartoon}. $k_1$, $k_2$, and $k_3$ (columns 3--5) denote the stellar types \citep[adopting the labeling scheme of \texttt{SSE};][]{Hurley2000} of the two collision inputs and the collision product, respectively. Columns 6--8 show the total stellar masses of the three stars and columns 9--11 show the core masses. Columns 12 and 13 show the impact parameter and velocity at infinity for the two colliding stars.}
\end{deluxetable}

\begin{deluxetable}{c|c|ccc|ccc|ccc|cc}
\tabletypesize{\scriptsize}
\tablewidth{0pt}
\tablecaption{Collision history for pair-instability BH; simulation \texttt{2c}.\label{table:collision_history2}}
\tablehead{
	\colhead{} &
	\colhead{Time} &
	\colhead{$k_1$} &
	\colhead{$k_2$} &
	\colhead{$k_3$} &
	\colhead{$M_1$} &
	\colhead{$M_2$} &
	\colhead{$M_3$} &
	\colhead{$M_{\rm{core},1}$} &
	\colhead{$M_{\rm{core},2}$} &
	\colhead{$M_{\rm{core},3}$} &	
    \colhead{$b$} &
	\colhead{$v_{\infty}$}\\
	\colhead{} &
	\colhead{(Myr)} &
	\colhead{} &
	\colhead{} &
	\colhead{} &
	\multicolumn{3}{c}{($M_{\odot}$)} &
	\multicolumn{3}{c}{($M_{\odot}$)} &
    \colhead{($R_{\odot}$)} &
	\colhead{($\rm{km\,s}^{-1}$)}
}
\startdata
1 & 3.228 & 1 & 1 & 1 & 0.9 & 88.0 & 89.2 & 0.0 & 0.0 & 0.0 & 254.6 & 76.1 \\
2 & 3.757 & 0 & 4 & 4 & 0.2 & 66.0 & 66.3 & 0.0 & 42.0 & 42.0 & 2216.0 & 68.0 \\
3 & 3.799 & 1 & 4 & 4 & 47.9 & 60.0 & 107.4 & 0.0 & 42.3 & 42.3 & 19816.0 & 24.2 \\
\hline
4 & 3.999 & - & - & 14 & - & - & 69.9 & & & & &  \\
\enddata
\tablecomments{Collision history for the $69.9\,M_{\odot}$ BH show in Figure \ref{fig:mass_vs_time}.}
\end{deluxetable}

\begin{deluxetable}{c|c|ccc|ccc|ccc|cc}
\tabletypesize{\scriptsize}
\tablewidth{0pt}
\tablecaption{PISN collision history; simulation \texttt{2c}.\label{table:collision_history3}}
\tablehead{
	\colhead{} &
	\colhead{Time} &
	\colhead{$k_1$} &
	\colhead{$k_2$} &
	\colhead{$k_3$} &
	\colhead{$M_1$} &
	\colhead{$M_2$} &
	\colhead{$M_3$} &
	\colhead{$M_{\rm{core},1}$} &
	\colhead{$M_{\rm{core},2}$} &
	\colhead{$M_{\rm{core},3}$} &	
    \colhead{$b$} &
	\colhead{$v_{\infty}$}\\
	\colhead{} &
	\colhead{(Myr)} &
	\colhead{} &
	\colhead{} &
	\colhead{} &
	\multicolumn{3}{c}{($M_{\odot}$)} &
	\multicolumn{3}{c}{($M_{\odot}$)} &
    \colhead{($R_{\odot}$)} &
	\colhead{($\rm{km\,s}^{-1}$)}
}
\startdata
1 & 2.059 & 1 & 1 & 1 & 66.8 & 120.1 & 186.9 & 0.0 & 0.0 & 0.0 & 812.4 & 43.2 \\
2 & 3.497 & 4 & 4 & 4 & 83.0 & 93.0 & 175.4 & 46.2 & 45.1 & 91.3 & 22794.8 & 33.8 \\
3 & 3.606 & 0 & 4 & 4 & 0.1 & 159.0 & 158.8 & 0.0 & 93.0 & 93.0 & 18220.9 & 75.0 \\
4 & 3.736 & 1 & 4 & 4 & 68.3 & 140.0 & 207.5 & 0.0 & 95.0 & 95.0 & 43961.7 & 24.7 \\
5 & 3.742 & 1 & 4 & 4 & 0.9 & 207.0 & 207.5 & 0.0 & 95.0 & 95.0 & 11824.2 & 67.9 \\
\hline
6 & 3.742 & - & - & 15 & - & - & 0.0 & & & & &  \\
\enddata
\tablecomments{Collision history for the PISN show in Figure \ref{fig:mass_vs_time}.}
\end{deluxetable}

\begin{deluxetable}{c|c|ccc|ccc|ccc|cc}
\tabletypesize{\scriptsize}
\tablewidth{0pt}
\tablecaption{Collision history for runaway in Simulation \texttt{1a}.\label{table:collision_history4}}
\tablehead{
	\colhead{} &
	\colhead{Time} &
	\colhead{$k_1$} &
	\colhead{$k_2$} &
	\colhead{$k_3$} &
	\colhead{$M_1$} &
	\colhead{$M_2$} &
	\colhead{$M_3$} &
	\colhead{$M_{\rm{core},1}$} &
	\colhead{$M_{\rm{core},2}$} &
	\colhead{$M_{\rm{core},3}$} &	
    \colhead{$b$} &
	\colhead{$v_{\infty}$}\\
	\colhead{} &
	\colhead{(Myr)} &
	\colhead{} &
	\colhead{} &
	\colhead{} &
	\multicolumn{3}{c}{($M_{\odot}$)} &
	\multicolumn{3}{c}{($M_{\odot}$)} &
    \colhead{($R_{\odot}$)} &
	\colhead{($\rm{km\,s}^{-1}$)}
}
\startdata
1 & 3.349 & 4 & 1 & 4 & 91.4 & 92.0 & 183.1 & 45.4 & 0.0 & 45.4 & 26727.3 & 25.9 \\
2 & 3.491 & 4 & 4 & 4 & 78.2 & 161.0 & 239.2 & 46.2 & 46.4 & 92.6 & 35007.0 & 31.1 \\
3 & 3.497 & 1 & 4 & 4 & 7.7 & 238.0 & 246.1 & 0.0 & 92.7 & 92.7 & 16795.6 & 71.8 \\
4 & 3.519 & 4 & 4 & 4 & 90.0 & 243.0 & 331.9 & 45.7 & 93.0 & 138.7 & 59561.7 & 21.4 \\
5 & 3.522 & 0 & 4 & 4 & 0.4 & 331.0 & 331.8 & 0.0 & 138.8 & 138.8 & 21606.0 & 68.0 \\
6 & 3.524 & 1 & 4 & 4 & 31.5 & 332.0 & 362.8 & 0.0 & 138.8 & 138.8 & 85056.5 & 14.8 \\
7 & 3.559 & 1 & 4 & 4 & 2.9 & 358.0 & 360.4 & 0.0 & 139.7 & 139.7 & 15062.1 & 72.7 \\
8 & 3.563 & 4 & 4 & 4 & 178.2 & 360.0 & 537.3 & 46.8 & 139.8 & 186.6 & 156524.0 & 13.0 \\
9 & 3.565 & 4 & 4 & 4 & 65.4 & 537.0 & 601.5 & 46.8 & 186.7 & 233.5 & 62288.0 & 37.0 \\
10 & 3.567 & 1 & 4 & 4 & 25.4 & 601.0 & 626.3 & 0.0 & 233.6 & 233.6 & 133810.0 & 15.2 \\
11 & 3.592 & 1 & 4 & 4 & 4.4 & 623.0 & 626.9 & 0.0 & 234.6 & 234.6 & 32677.6 & 80.6 \\
12 & 3.593 & 1 & 4 & 4 & 4.9 & 627.0 & 631.5 & 0.0 & 234.6 & 234.6 & 61943.9 & 41.2 \\
13 & 3.613 & 4 & 4 & 4 & 80.8 & 628.0 & 708.4 & 41.6 & 235.5 & 277.1 & 73274.9 & 23.6 \\
14 & 3.614 & 0 & 4 & 4 & 0.2 & 708.0 & 708.4 & 0.0 & 277.1 & 277.1 & 29014.8 & 82.2 \\
15 & 3.619 & 1 & 4 & 4 & 25.3 & 708.0 & 732.7 & 0.0 & 277.3 & 277.3 & 35068.2 & 23.9 \\
16 & 3.62 & 4 & 4 & 4 & 695.7 & 733.0 & 1426.8 & 139.3 & 277.4 & 416.7 & 224846.0 & 11.6 \\
\enddata
\end{deluxetable}

\begin{deluxetable}{c|c|ccc|ccc|ccc|cc}
\tabletypesize{\scriptsize}
\tablewidth{0pt}
\tablecaption{Collision history for runaway in Simulation \texttt{1b}. \label{table:collision_history5}}
\tablehead{
	\colhead{} &
	\colhead{Time} &
	\colhead{$k_1$} &
	\colhead{$k_2$} &
	\colhead{$k_3$} &
	\colhead{$M_1$} &
	\colhead{$M_2$} &
	\colhead{$M_3$} &
	\colhead{$M_{\rm{core},1}$} &
	\colhead{$M_{\rm{core},2}$} &
	\colhead{$M_{\rm{core},3}$} &	
    \colhead{$b$} &
	\colhead{$v_{\infty}$}\\
	\colhead{} &
	\colhead{(Myr)} &
	\colhead{} &
	\colhead{} &
	\colhead{} &
	\multicolumn{3}{c}{($M_{\odot}$)} &
	\multicolumn{3}{c}{($M_{\odot}$)} &
    \colhead{($R_{\odot}$)} &
	\colhead{($\rm{km\,s}^{-1}$)}
}
\startdata
1 & 3.352 & 4 & 4 & 4 & 92.8 & 95.0 & 187.2 & 45.5 & 45.4 & 90.9 & 22096.2 & 49.4 \\
2 & 3.41 & 1 & 4 & 4 & 47.8 & 178.0 & 225.7 & 0.0 & 91.8 & 91.8 & 50954.1 & 17.3 \\
3 & 3.472 & 4 & 4 & 4 & 123.1 & 216.0 & 338.5 & 46.6 & 92.8 & 139.4 & 56550.8 & 24.5 \\
4 & 3.5 & 1 & 4 & 4 & 7.3 & 334.0 & 341.5 & 0.0 & 140.0 & 140.0 & 20029.7 & 77.0 \\
5 & 3.527 & 1 & 4 & 4 & 3.2 & 337.0 & 340.6 & 0.0 & 140.7 & 140.7 & 59184.8 & 18.9 \\
6 & 3.535 & 1 & 4 & 4 & 87.4 & 339.0 & 426.5 & 0.0 & 140.9 & 140.9 & 4208.3 & 73.6 \\
7 & 3.541 & 1 & 4 & 4 & 2.4 & 426.0 & 428.0 & 0.0 & 141.0 & 141.0 & 44744.4 & 26.2 \\
8 & 3.545 & 1 & 4 & 4 & 13.6 & 427.0 & 440.9 & 0.0 & 141.1 & 141.1 & 70184.8 & 18.3 \\
9 & 3.548 & 1 & 4 & 4 & 56.7 & 440.0 & 497.0 & 0.0 & 141.2 & 141.2 & 18116.4 & 68.3 \\
10 & 3.569 & 4 & 4 & 4 & 444.6 & 494.0 & 936.5 & 185.9 & 141.7 & 327.6 & 37210.6 & 19.0 \\
11 & 3.571 & 1 & 4 & 4 & 59.1 & 936.0 & 995.1 & 0.0 & 327.8 & 327.8 & 150765.0 & 15.1 \\
12 & 3.574 & 1 & 4 & 4 & 21.2 & 995.0 & 1015.7 & 0.0 & 327.9 & 327.9 & 190181.0 & 16.1 \\
\enddata
\end{deluxetable}

\begin{deluxetable}{c|c|ccc|ccc|ccc|cc}
\tabletypesize{\scriptsize}
\tablewidth{0pt}
\tablecaption{Collision history for runaway in Simulation \texttt{1c}. \label{table:collision_history6}}
\tablehead{
	\colhead{} &
	\colhead{Time} &
	\colhead{$k_1$} &
	\colhead{$k_2$} &
	\colhead{$k_3$} &
	\colhead{$M_1$} &
	\colhead{$M_2$} &
	\colhead{$M_3$} &
	\colhead{$M_{\rm{core},1}$} &
	\colhead{$M_{\rm{core},2}$} &
	\colhead{$M_{\rm{core},3}$} &	
    \colhead{$b$} &
	\colhead{$v_{\infty}$}\\
	\colhead{} &
	\colhead{(Myr)} &
	\colhead{} &
	\colhead{} &
	\colhead{} &
	\multicolumn{3}{c}{($M_{\odot}$)} &
	\multicolumn{3}{c}{($M_{\odot}$)} &
    \colhead{($R_{\odot}$)} &
	\colhead{($\rm{km\,s}^{-1}$)}
}
\startdata
1 & 3.446 & 4 & 4 & 4 & 82.3 & 93.0 & 174.3 & 45.6 & 45.3 & 90.9 & 27762.1 & 29.3 \\
2 & 3.488 & 1 & 4 & 4 & 3.2 & 168.0 & 171.2 & 0.0 & 91.6 & 91.6 & 9218.1 & 81.3 \\
3 & 3.546 & 1 & 4 & 4 & 59.6 & 162.0 & 221.6 & 0.0 & 92.5 & 92.5 & 36207.6 & 21.8 \\
4 & 3.572 & 1 & 4 & 4 & 52.8 & 218.0 & 270.2 & 0.0 & 92.9 & 92.9 & 22892.2 & 49.4 \\
5 & 3.59 & 4 & 4 & 4 & 85.1 & 268.0 & 352.0 & 41.0 & 93.1 & 134.2 & 117451.0 & 13.3 \\
6 & 3.622 & 1 & 4 & 4 & 8.4 & 347.0 & 355.4 & 0.0 & 134.9 & 134.9 & 19602.9 & 50.3 \\
7 & 3.702 & 4 & 4 & 4 & 147.6 & 343.0 & 490.4 & 40.9 & 136.8 & 177.6 & 38033.3 & 37.3 \\
8 & 3.704 & 4 & 4 & 4 & 363.3 & 490.0 & 851.0 & 143.2 & 177.7 & 320.9 & 102791.0 & 34.0 \\
9 & 3.706 & 1 & 4 & 4 & 57.4 & 851.0 & 907.7 & 0.0 & 321.1 & 321.1 & 39229.5 & 37.6 \\
10 & 3.709 & 1 & 4 & 4 & 19.6 & 907.0 & 926.7 & 0.0 & 321.3 & 321.3 & 837690.0 & 3.8 \\
11 & 3.71 & 4 & 4 & 4 & 550.5 & 927.0 & 1474.9 & 144.2 & 321.3 & 465.5 & 380430.0 & 6.7 \\
\enddata
\end{deluxetable}

\end{document}